\documentclass[a4paper,fleqn,usenatbib]{mnras}

\usepackage{newtxtext,newtxmath}
\usepackage[T1]{fontenc}
\usepackage{ae,aecompl}

\usepackage{graphicx}	% Including figure files
\usepackage{amsmath}	% Advanced maths commands
\usepackage{amssymb}	% Extra maths symbols
\usepackage{comment}	% Extra maths symbols
\usepackage{color}	% Extra maths symbols
\usepackage{tcolorbox}
\usepackage[export]{adjustbox}

\title[Remnant FRII Radio Galaxies]{On the population of remnant FRII radio galaxies and implications for radio source dynamics.}

\author[L. E.H. Godfrey et al.]{
L. E.H. Godfrey,$^{1}$\thanks{E-mail: leith.godfrey@gmail.com}
R. Morganti,$^{1, 2}$
M. Brienza,$^{1, 2}$
\\
$^{1}$ASTRON, the Netherlands Institute for Radio Astronomy, Postbus 2, 7990 AA, Dwingeloo, The Netherlands. \\ 
$^{2}$Kapteyn Astronomical Institute, University of Groningen, Postbus 800, 9700 AV Groningen, The Netherlands.
}

\pubyear{2017}

\begin{document}
\label{firstpage}
\pagerange{\pageref{firstpage}--\pageref{lastpage}}
\maketitle

\begin{abstract}

The purpose of this work is two-fold: (1)  to quantify the occurrence of ultra-steep spectrum remnant FRII radio galaxies in a 74 MHz flux limited sample, and (2) perform Monte-Carlo simulations of the population of active and remnant FRII radio galaxies to confront models of remnant lobe evolution, and provide guidance for further investigation of remnant radio galaxies. We find that fewer than 2$\%$ of FRII radio galaxies with S$_{ \rm74~MHz} > 1.5$~Jy are candidate ultra-steep spectrum remnants, where we define ultra-steep spectrum as $\alpha_{\rm 74~MHz}^{\rm 1400~MHz} > 1.2$. Our Monte-Carlo simulations demonstrate that models involving Sedov-like expansion in the remnant phase, resulting in rapid adiabatic energy losses, are consistent with this upper limit, and predict the existence of nearly twice as many remnants with normal (not ultra-steep) spectra in the observed frequency range as there are ultra-steep spectrum remnants. This model also predicts an ultra-steep remnant fraction approaching 10$\%$ at redshifts $z < 0.5$. Importantly, this model implies the lobes remain over-pressured with respect to the ambient medium well after their active lifetime, in contrast with existing observational evidence that many FRII radio galaxy lobes reach pressure equilibrium with the external medium whilst still in the active phase.  The predicted age distribution of remnants is a steeply decreasing function of age. In other words young remnants are expected to be much more common than old remnants in flux limited samples.  For this reason, incorporating higher frequency data $\gtrsim 5$~GHz will be of great benefit to future studies of the remnant population.  

\end{abstract}


%%%%%%%%%%%%%%%%% BODY OF PAPER %%%%%%%%%%%%%%%%%%

\section{Introduction}  \label{sec:introduction}

%@arxiver{remnant_fraction_vs_redshift_sedov.pdf,histograms_sedov.pdf}

Do Fanaroff-Riley type II (FRII; Fanaroff \& Riley, 1974) radio galaxies remain over-pressured with respect to the external medium throughout their entire active life, or do they at some point reach pressure equilibrium with their environment? The answer to this question has potentially far-reaching implications for the understanding of radio galaxy dynamics and energetics. If the lobes remain over-pressured throughout the entire life, the source is always surrounded by an elliptical bow shock and the lobes undergo supersonic self-similar expansion \citep{ka97}. This self-similar scenario is typically assumed when modelling populations of FRII radio galaxies \citep[for example][to name a few]{blundell99, wang08, kapinska12}. However, some studies now point to a more complex situation, in which the lobes start out being highly over-pressured, but reach pressure equilibrium with their environment well before the jet activity comes to an end. Estimates of internal lobe pressures based on the assumption of minimum energy \citep{hardcastle00} as well as those based on inverse-Compton measurements \citep{hardcastle02, croston04} are observed to be comparable to the external pressures, assuming no large contribution to the lobe pressure from thermal material within the lobes. Furthermore, \citet{mullin08} find a trend of increasing axial ratio with sources size in a flux limited sample of FRIIs with $z<1$. If the lobes remained over-pressured throughout their lifetime, undergoing self-similar expansion, then the axial ratio is expected to remain constant. The observed size-dependent axial ratio distribution demonstrated by \citet{mullin08} is therefore inconsistent with models in which the expansion remains self-similar. \citet{hardcastle00} and \citet{mullin08} suggest that FRII radio galaxies may only grow self-similarly early on in their lifetime, and reach pressure equilibrium with their surroundings in middle-age. 

One way to discriminate between these two scenarios is to consider the remnant phase of radio galaxy evolution. If the lobes remain highly over pressured throughout their lifetime, then the remnant phase will be governed by supersonic Sedov-like expansion \citep{kaiser02}. This continued supersonic lobe expansion in the remnant phase will cause rapid dimming of the lobe radio emission due to a decrease in the magnetic field strength and decrease in particle energies due to adiabatic expansion losses. In contrast, if the lobes are already in pressure equilibrium with the external medium at the end of the active phase, the luminosity evolution in the remnant phase is expected to be much more sedate. 

\citet{kaiser02} showed that models of remnant radio galaxy spectra are highly degenerate, and modelling of individual remnant radio galaxy spectra cannot constrain the history of lobe evolution. The only way to constrain the lobe evolution in the remnant phase is via a statistical approach. That is the approach taken in this paper. We compose a flux-limited sample radio sources that is dominated by FRII radio galaxies (Section \ref{sec:sample_selection}). With this sample, we obtain an upper limit on the number of ultra-steep spectrum remnant radio galaxies in a low frequency selected, flux limited sample of FRII radio galaxies (Section \ref{sec:empirical_results}). We then perform Monte-Carlo simulations to assess whether models of remnant phase lobe evolution are consistent with the observed limit on ultra-steep spectrum remnants in our flux limited sample (Sections \ref{sec:simulations} and \ref{sec:modeling_results}). 

We make a strong distinction between the population of ultra-steep spectrum remnants, which we define as remnant radio galaxies with spectral index $\alpha > 1.2$ (defined such that $S_\nu \propto \nu^{-\alpha}$) between our chosen frequencies, and the entire population of remnant radio galaxies, which includes any remnant radio galaxy regardless of spectral characteristics. This distinction is necessary because not all remnant radio galaxies have ultra-steep spectra in the observed frequency range. A good example of this is the remnant radio galaxy discovered by \citet{brienza16}, which was identified purely based on morphological characteristics, and only shows an ultra-steep spectrum above 1.4 GHz.  \\

In this work, we restrict our analysis to FRII radio galaxies. A follow-up paper (Brienza et al., 2017 submitted) will focus on the study of lower luminosity, FRI class remnant radio galaxies.

\section{Sample selection}  \label{sec:sample_selection}
% ==================================================================
% Sample selected using the following python script:
% /Users/godfrey/astro/Projects/Current_Focus/1_Remnant_RGs_and_Generalised_CI_model/VLSSr_x_NVSS_Sample_Selection/Get_VLSSxNVSS_sample.py
% ==================================================================

\subsection{Sample Definition}  \label{sec:VLSSr_sample_definition}

Our sample selection is based on the 74 MHz VLA Low-freqeuncy Sky Survey Redux catalog \citep[VLSSr;][]{lane14}. We calculate the flux density of sources in the VLSSr catalogue from the catalogued peak intensity, and fitted major and minor axes, based on the expressions given in \citet{condon97, cohen07}. We then restrict the VLSSr catalog as follows:
\begin{enumerate}
\item 9 hrs $<$ RA $<$ 16 hrs
\item $0^{\circ} <$ DEC $< 60^{\circ}$
\item Distance to nearest neighbour D $>$ 4 arcminutes.
\item Fitted major axis $<$ 120 arcseconds. 
\item Flux density $S_{\rm 74~MHz} > 1.5$~Jy
\end{enumerate}
The reason for each of the restrictions is as follows: 

(i and ii) The restrictions on RA and DEC are imposed in order to match the sky area of the FIRST survey, so that we can assess the morphology of the selected objects. 

(iii) The median distance to the nearest neighbour in the VLSSr catalogue is approximately 15 arcminutes. However, a histogram of distance to nearest neighbour shows a clear peak in the range 1 - 4 arcminutes. The narrow peak at 1-4 arcminutes is a result of radio galaxies with large angular size and complex morphology, in which the individual radio galaxies are fitted by more than one Gaussian component. To simplify the cross-matching with NVSS, we restrict our sample to ``isolated" VLSSr sources, for which the distance to nearest neighbour is greater than 4 arcminutes. This selection criterion reduces the catalogue size by 8$\%$. However, for sources with nearest neighbour $<$ 4 arcminutes, several catalogued sources are often related to the same radio galaxy, and so the reduction in the number of radio galaxies is likely to be significantly less than 8$\%$. 

(iv) We next remove all objects for which the fitted major axis size is equal to 120 arcseconds, which is the upper limit on fitted major axis \citep{lane14}. For these objects, we are unable to accurately calculate the flux density based on the catalogued values of peak intensity, major and minor axes. This selection criterion reduces the catalogue size by a further 8$\%$ (1237 objects). These very large angular size sources would provide an interesting sample for searches for remnant radio galaxies, as they most likely represent very large, relatively nearby radio galaxies. However, the need to obtain accurate flux density estimates from the catalogued values of peak intensity, major axis and minor axis, necessitates the removal of these large sources from our analysis. 

(v) Finally, we impose a flux density limit of $S_{\rm 74~MHz} > 1.5$~Jy, which corresponds to the knee in the source-counts distribution \citep[eg.][]{massardi10}. With this flux limit, the sample is expected to be highly dominated by high excitation (predominantly FRII) radio galaxies, with an FRII fraction of up to 80$\%$ \citep{willott01}. We note that the S-cubed simulation of \citet{wilman08} predicts that our sample will comprise 55$\%$ FRII, 5$\%$ Gigahertz Peaked Spectrum (GPS), and 40$\%$ FRI\footnote{The lowest frequency in the S-cubed database is 151 MHz, so we converted our 74 MHz flux limit to 151 MHz assuming a spectral index of 0.75.}. The morphologies of our sample 
%(Figure \ref{fig:first_morphologies_entire_sample}), 
agree with the prediction of the S-cubed simulation: $\gtrsim 50\%$ of our sample have a double-lobed FRII-like appearance. Whilst the sample contains only $\gtrsim 50\%$ FRII radio galaxies, we account for this factor in our analysis of the FRII remnant population, as described in section \ref{sec:FIRST_morphologies_entire_sample}.

The flux limit further reduces the sample size by 72$\%$, giving a final sample size of 3861 objects. 

\subsection{Cross-matching with NVSS}

We cross-match our 74MHz flux limited sample described in Section \ref{sec:VLSSr_sample_definition} with the NVSS catalogue at 1.4 GHz \citep{condon98}. Due to the higher frequency and higher resolution of the NVSS relative to the VLSSr, individual catalogue entries in the VLSSr may be associated with multiple catalogue entries in the NVSS. For this reason, we must use a large matching radius. Given that the largest fitted major axis in the VLSSr sample is 120 arcseconds, we use a matching radius of 60 arcseconds, and sum the flux densities of all NVSS catalogue entries within the matching radius. The results of our cross-matching is summarised in Table \ref{table:cross_match_results}.

\begin{figure}
%	\includegraphics[width=2.0\columnwidth]{figures/2D_panels_plot_327MHz.pdf}
   	\includegraphics[width=1.0\columnwidth]{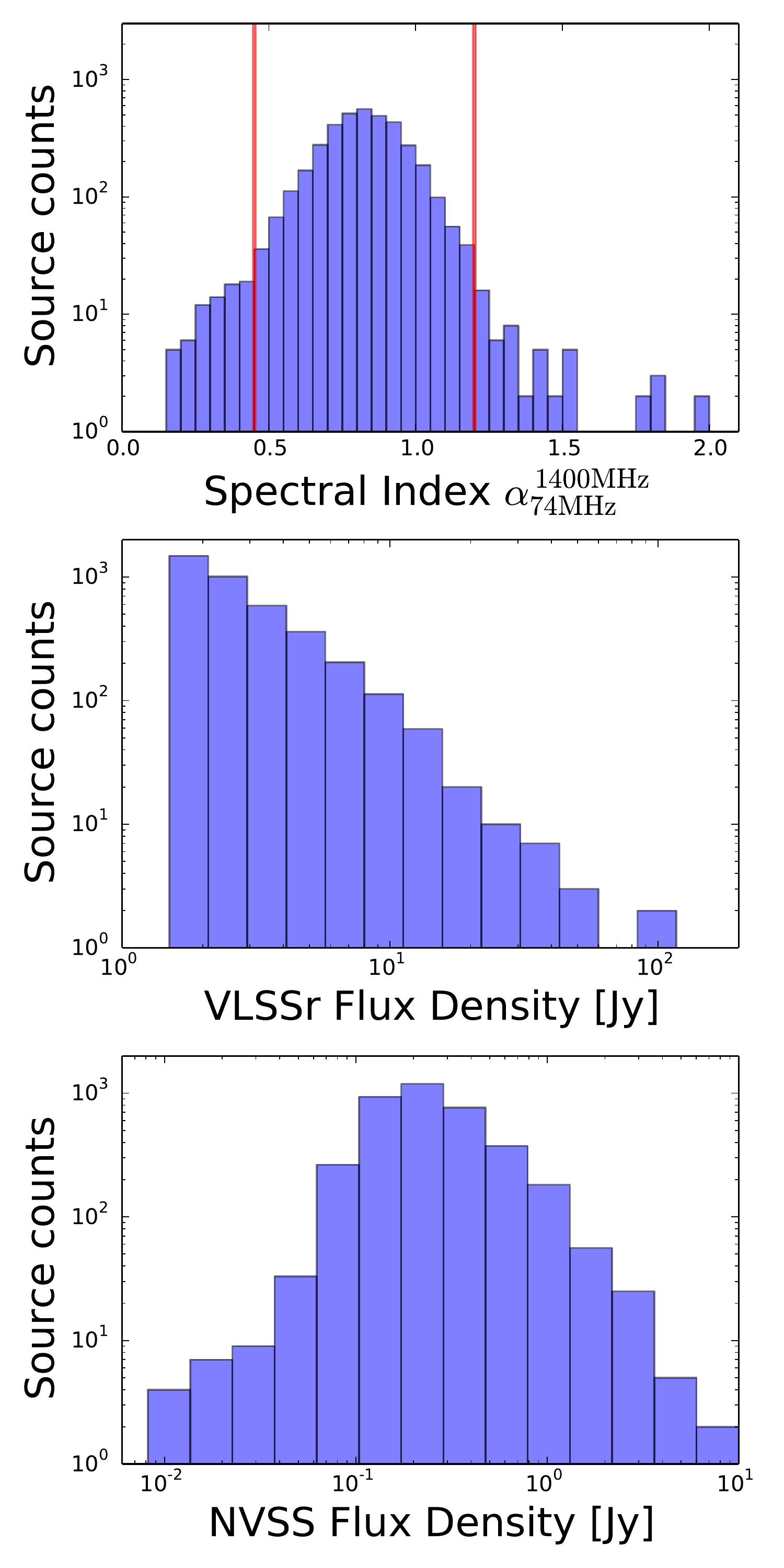}
    \caption{Histograms of spectral index and flux densities for our VLSSr selected sample. The histogram of spectral index clearly consists of a broad central peak, along with a flat spectrum tail and ultra-steep spectrum tail. The two vertical lines in the upper panel denote our division of the sample into flat spectrum ($\alpha < 0.45$), normal spectrum (0.45 < $\alpha < 1.2$), and ultra-steep spectrum ($\alpha > 1.2$). Note that 4 objects are not detected in NVSS, and are not included in the above histograms.}
    \label{fig:spectral_index_distribution}
 %-----
 % This plot produced by /Users/godfrey/astro/Projects/Current_Focus/1_Remnant_RGs_and_Generalised_CI_model/VLSSr_x_NVSS_Sample_Selection/Get_VLSSxNVSS_sample.py 
 %-----
 \end{figure}

The probability of finding an NVSS source within 60 arcseconds of a random position is approximately 4$\%$ \citep{condon98}. We therefore expect that approximately 4$\%$ of our sample ($\sim 150$ objects) will contain an unrelated NVSS source within the search radius of 60 arcseconds. However, these unrelated sources are clustered near the NVSS flux limit (2.5 mJy), while the majority of our sample have NVSS flux densities more than 10 times higher than the flux limit (see Figure \ref{fig:spectral_index_distribution}). Therefore, only a very small fraction of our sample will be significantly affected by the detection of an unrelated NVSS object within the search radius.

The differing survey resolution is not likely to cause significant errors in spectral index calculation. Our sample is restricted to sources with fitted major axes less than 120 arcseconds in the VLSSr catalogue, a factor of only 2.7 times the NVSS angular resolution of 45 arcseconds. Furthermore, our sample has a high signal to noise in the NVSS survey, and therefore the catalogued flux densities are unlikely to have missed flux.

\begin{table}
 \caption{Results of cross-matching our VLSSr selected sample with NVSS.}
 \label{table:cross_match_results}
  \begin{tabular}{cc}
  \hline
Number of NVSS matches & Number of VLSSr sources  \\
  \hline
0 & 4 \\
1 & 3466 \\
2 & 382  \\
3 & 9  \\
$>3$ & 0  \\
\hline
Total & 3861 \\
  \hline
\end{tabular}
\end{table}

\begin{table*}
 \caption{Results of cross-matching our VLSSr sample with the FRII candidate sample of \citet{vanvelzen15}.}
 \label{table:cross_match_results_dubbeltjes}
  \begin{tabular}{ccc}
  \hline
  & Entire VLSSr selected Sample & Ultra-steep Spectrum Sample  \\
  & & ($\alpha_{\rm VLSSr}^{\rm NVSS} > 1.2$)  \\
  \hline
Sample Size & 3861 & 57  \\
Fraction matched with FRII candidates & 65$\%$ (2498/3861) & 45$\%$ (26/57)  \\
Fraction of FRII candidates with detected core & 14$\%$ (348/2498) & 23$\%$ (6/26) \\
  \hline
\end{tabular}
\end{table*}

\section{Empirical Results}  \label{sec:empirical_results}

In Figure \ref{fig:spectral_index_distribution} we present the histogram of spectral index between 74 MHz and 1400 MHz for our sample, as well as the histogram of 74 MHz and 1400 MHz flux densities. The spectral index distribution shows a broad, symmetric central peak, along with a flat spectrum tail and an ultra-steep spectrum tail. We wish to place an upper limit on the fraction of our sample that are ultra-steep spectrum remnant radio galaxies. To do so, we split our sample into 3 spectral categories: flat ($\alpha < 0.45$); normal ($0.45 < \alpha < 1.2$); and ultra-steep ($\alpha > 1.2$). The choice of the dividing line between flat, normal and ultra-steep spectrum is somewhat arbitrary, but is guided by the shape of the histogram in Figure \ref{fig:spectral_index_distribution}, and models of radio source emission. The ultra-steep spectrum limit at $\alpha = 1.2$ corresponds to the maximum spectral index for active radio galaxy models with particle injection corresponding to $\alpha < 0.7$ and a cooling break of $\Delta \alpha = 0.5$. We stress that our choice of dividing line at $\alpha = 1.2$ is not intended to capture all remnant radio galaxies. Indeed, we expect that most remnant radio galaxies above our flux limit will not have ultra-steep spectra according to our definition. However, the key results are not strongly affected by our choice of dividing line, since our modelling approach is designed specifically to accommodate arbitrary selection criteria. 

Flat spectrum objects comprise $2\%$ of the sample (78 objects), ``normal" spectrum objects comprise 96.5$\%$ of our sample (3726 objects), and ultra-steep spectrum objects comprise only $1.5\%$ of the sample (57 objects). Note, however, that our sample is comprised of both FRI and FRII radio galaxies. In this work, we seek an upper limit on the fraction of \emph{FRII radio galaxies} that are ultra-steep spectrum remnants. We therefore need to account for the fraction of the sample that is FRII. We consider this question in the following sections.

\begin{figure*}
%	\includegraphics[width=2.0\columnwidth]{figures/2D_panels_plot_327MHz.pdf}
   	\includegraphics[width=2.0\columnwidth]{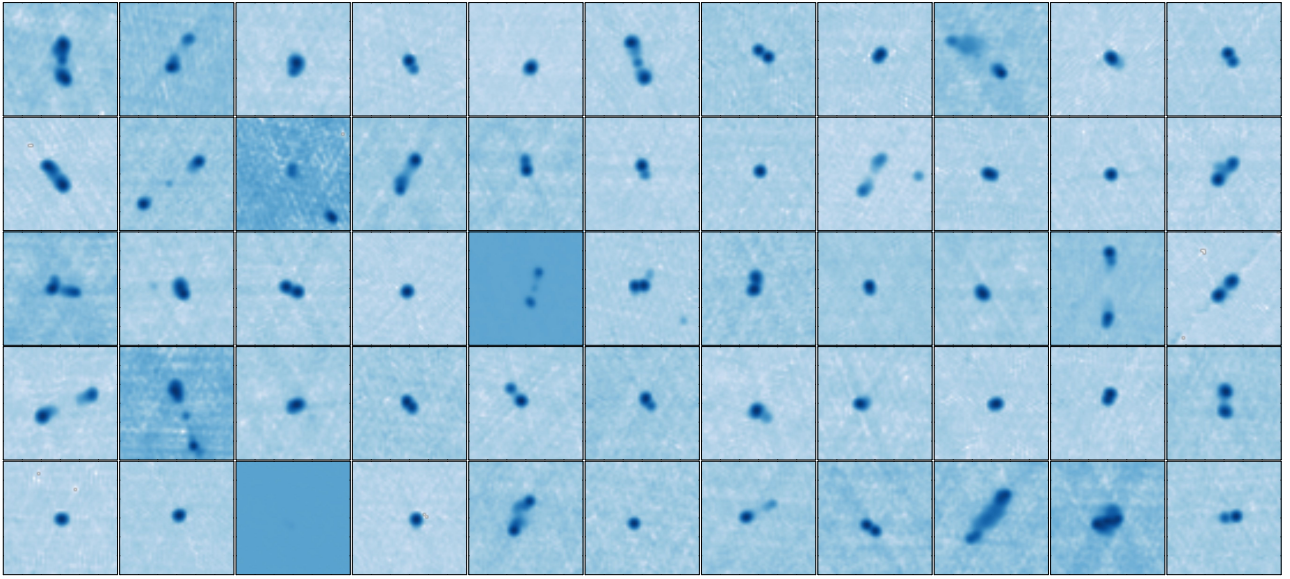}
    \caption{1.4 GHz radio images from the FIRST survey of 55 randomly selected sources in our VLSSr selected sample with normal spectral index values ($0.45 < \alpha_{\rm VLSS}^{\rm NVSS} < 1.2$). The images are 2 x 2 arcminutes.}
    \label{fig:first_morphologies_entire_sample}
 %-----
 % This plot produced by /Users/godfrey/astro/Projects/Current_Focus/1_Remnant_RGs_and_Generalised_CI_model/VLSSr_x_NVSS_Sample_Selection/Get_VLSSxNVSS_sample.py 
 %-----
 \end{figure*}

\begin{figure*}
%	\includegraphics[width=2.0\columnwidth]{figures/2D_panels_plot_327MHz.pdf}
   	\includegraphics[width=2.0\columnwidth]{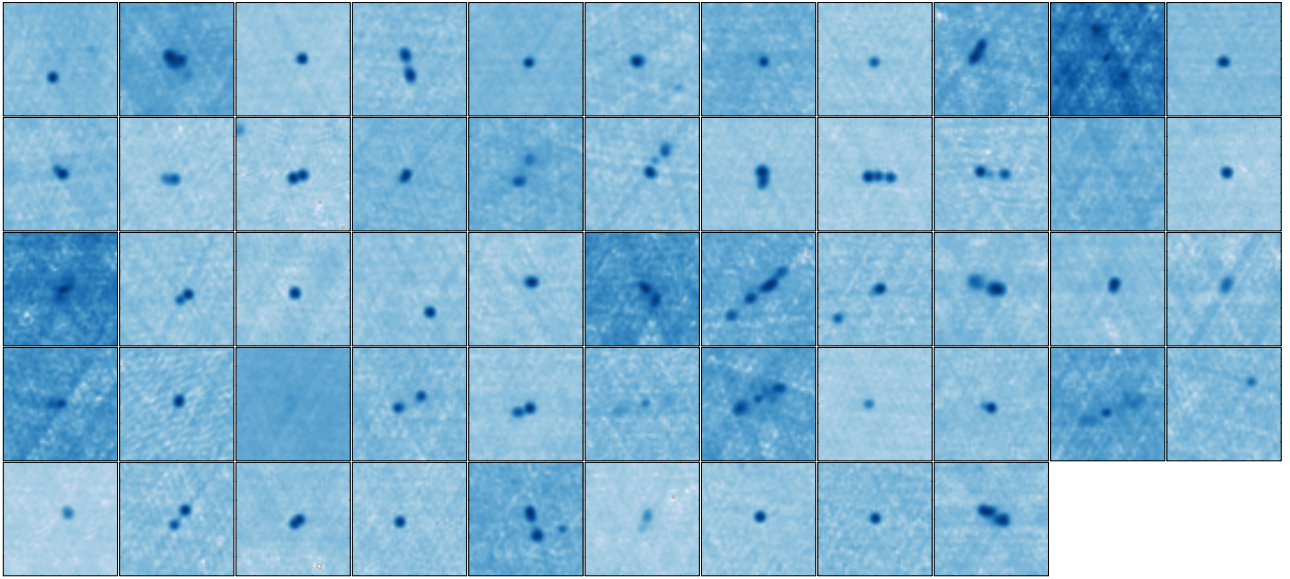}
    \caption{1.4 GHz radio images from the FIRST survey of the 53 VLSSr selected sources detected in NVSS with ultra-steep spectra ($\alpha_{\rm VLSS}^{\rm NVSS} > 1.2$). Many of the ultra-steep spectrum sources show evidence of active cores, and may represent re-started radio galaxies. Many are unresolved, and may be associated with variable AGN cores. The images are 2 x 2 arcminutes.}
    \label{fig:first_morphologies_steep_spectrum_sample}
 %-----
 % This plot produced by /Users/godfrey/astro/Projects/Current_Focus/1_Remnant_RGs_and_Generalised_CI_model/VLSSr_x_NVSS_Sample_Selection/Get_VLSSxNVSS_sample.py 
 %-----
 \end{figure*}

\subsection{Morphology of the S$_{\rm 74~MHz} >$ 1.5~Jy sample in FIRST}  \label{sec:FIRST_morphologies_entire_sample}

Our flux density limit of 1.5 Jy at 74 MHz corresponds to the knee in the source counts distribution. This flux limit is chosen to provide a large sample that is dominated by high excitation (predominantly FRII) radio galaxies. 

To assess the dominant morphology of the sample, we compare our sample to that of \citet{vanvelzen15}, who compiled a sample of 59,192 candidate FRII radio sources from the FIRST survey catalogue, using an algorithm that selected double, triple, or multiple sources with lobe-lobe separation of less than 60 arcseconds. This sample of ``dubbeltjes" (small doubles) is considered to be a relatively clean sample of FRII radio galaxies on these angular scales. We cross-match our VLSSr flux limited sample with the dubbeltjes sample of \citet{vanvelzen15}, using a matching radius of 30 arcseconds ($\lesssim 5$ expected random associations), and find 65$\%$ of our sample (2498 objects) are matched to candidate FRII radio galaxies in the catalogue of \citet{vanvelzen15}. Not all of the \citet{vanvelzen15} sample are confirmed FRIIs. However, the selection of FRIIs in \citet{vanvelzen15} only includes radio galaxies up to 1 arcminute in angular size. Our sample includes sources with de-convolved major axes up to approximately 100 arcseconds. We therefore visually inspected a random sample of FIRST images centred on the VLSSr coordinates, and find that more than 50$\%$ have a double-lobed FRII-like appearance, while 10$\%$ are unresolved at the FIRST resolution (see Figure \ref{fig:first_morphologies_entire_sample}). This random sample is drawn from our VLSSr flux limited sample, and includes sources that are within the van Velzen sample, as well as some that are not in that sample. 

Based on the preceding discussion, we can be confident that our $S_{\rm 74~ MHz} > 1.5$~Jy sample is dominated by FRII radio galaxies, as expected. We therefore place a robust lower limit on the number of FRII radio galaxies in our sample: $N_{\rm FRII} > 0.5 \times 3861 > 1930$ objects. 

\subsection{A limit on the fraction of FRIIs in our sample that are ultra-steep spectrum remnants}  \label{sec:empirical_limit_on_remnant_fraction}

The total number of ultra-steep spectrum ($\alpha_{\rm 74~MHz}^{\rm 1400~MHz} > 1.2$) sources in our sample is 57 (53 detections + 4 non-detections in NVSS). Therefore, as an absolute upper limit, fewer than 3$\%$ (57/1930) of FRII radio galaxies in our flux limited sample are ultra-steep spectrum remnants. 

Of course, not all ultra-steep spectrum radio sources are remnant radio galaxies, and therefore we can place a more stringent upper limit on the fraction of remnant FRIIs in our sample by considering what fraction of the ultra-steep spectrum objects in our sample are active. We do this initially by considering the ultra-steep spectrum sample object's morphology in FIRST. 
Of the 53 ultra-steep spectrum sources detected in NVSS, 10 are point-like in FIRST and may be high redshift radio galaxies, or core dominated systems in which the spectral index has been influenced by variable flux between the dates of observation in VLSS and NVSS surveys. A further 6 objects are triple systems, exhibiting two lobes plus an AGN core, indicating that these systems are still active (although some may be re-started systems, with remnant lobes and a re-born AGN). After removing the point-like and triple systems from the ultra-steep spectrum sample, we find that less than 41 objects are remnant radio galaxies, indicating that less than 2$\%$ of the FRIIs in our sample are candidate remnant radio galaxies. 

The large core-detection fraction in the ultra-steep spectrum sample suggests that many of the ultra-steep spectrum objects without a core detection would have a detected core if we were to follow-up with sensitive high resolution observations. Therefore, the true number of ultra-steep spectrum FRII remnants in our sample is likely to be less than 41, because we detect such a large fraction of triples in our ultra-steep spectrum sample. Out of the 2498 FRII candidates from \citet{vanvelzen15} in our VLSSr sample, only 14$\%$ have a detected core. In comparison, of the 26 FRII candidates from \citet{vanvelzen15} in our ultra-steep spectrum sample, 23$\%$ have a detected core. Extrapolating from the 23$\%$ of ultra-steep sources with detected cores, it is conceivable that most of the ultra-steep spectrum objects in this flux limited sample are active, and would be found to have core emission if they were followed up with more sensitive high resolution observations. We note that the higher core fraction in the ultra-steep spectrum dubbeltjes from \citep{vanvelzen15} could arise if the ultra-steep spectrum dubbeltjes tend to be larger angular size objects, enabling core detection in a larger number of cases.

Some of the ultra-steep spectrum triple sources may be restarted objects (eg. double-doubles) or high redshift radio galaxies, and a follow-up study of this sample of ultra-steep spectrum triples is warranted, particularly given that they are significantly lower angular size objects than most known double-doubles. 

In summary, fewer than 2$\%$ of the FRII radio galaxies in our 74~MHz selected flux limited sample are candidate ultra-steep spectrum remnants. Due to the large core-detection fraction in our ultra-steep spectrum sample, we expect that many of the canidate remnants are in fact active, so that the true ultra-steep remnant fraction is likely to be significantly less than 2$\%$. Follow-up radio observations at higher resolution are required to confirm the nature of the ultra-steep spectrum objects, and which, if any, are indeed remnant FRII radio galaxies.

\section{Simulating the population of active and remnant FRII radio galaxies}  \label{sec:simulations}

The second part of our study involves simulating the flux limited sample of FRII radio galaxies, including both the active and remnant phase, using an accurate model of spectral evolution. This Monte-Carlo approach allows us to compare the models with the empirical results using the exact same selection criteria. It also allows us, for the benefit of future studies, to investigate the efficiency of alternative selection criteria including such things as spectral curvature, high frequency spectra, and redshift limits. The degeneracy involved in modelling the radio spectrum of an individual remnant radio galaxy at one point in it's life \citep{kaiser02} can be broken by observing many sources at different stages of their life, and modelling their spectral distributions.

In this section we present the relevant model equations used to construct the time dependent radio galaxy spectra, and we describe the procedure used to generate the simulated catalogues. 

\subsection{Flux density calculation assuming non-uniform magnetic field strength}

In order to accurately simulate a population of radio galaxies including remnants, we must accurately trace the spectral evolution in both the active and remnant phases. \citet{tribble91, tribble93} showed that a non-uniform distribution of magnetic field strengths gives rise to aged synchrotron spectra that differ significantly from those obtained under the assumption of a uniform magnetic field strength. We therefore follow \citet{tribble91, tribble93} and \citet{hardcastle13a}, and calculate the synchrotron spectra assuming that within each volume element of the lobes the magnetic field is distributed according to a Guassian-random field. That is, at each point within a volume element, the cartesian components of the field are each drawn independently from a Gaussian distribution with mean of zero. This magnetic field configuration is expected to arise from homogenous, isotropic turbulence \citep{tribble91, hardcastle13a}. We further assume that each volume element within the lobes has the same magnetic field distribution, that the electron energy distribution is independent of the local magnetic field strength, and that the particle pitch angle distribution is isotropic. With these assumptions, we can calculate the synchrotron flux density as
\begin{equation} \label{eqn:S_nu}
S_\nu = \frac{(1+z)}{D_L^2} \frac{\sqrt{3} e^3 }{16 \pi^2 \epsilon_0 c m_e} \int_{\gamma_{\rm min}}^{\gamma_{\rm max}} N(\gamma) \left[ \int_0^\infty   B p_B \bar{F} (y) dB \right]d\gamma
\end{equation}
where $N(\gamma) = \frac{dN}{d\gamma}$ is the volume integrated electron energy distribution, $\bar{F}(y)$ is the angle-averaged synchrotron function (see Section \ref{app:F_bar_y}), $p_B$ is the probability distribution for the magnetic field strength, $\epsilon_0$ is the permittivity of free space, $c$ is the speed of light, $m_e$ is the electron mass, and $e$ is the electron charge. 

For a Guassian-random field, the probability distribution for the magnetic field strength $p_B$ is the Maxwell-Boltzmann distribution \citep[][]{hardcastle13a}:
\begin{eqnarray}
p_B = \sqrt{\frac{2}{\pi}} \frac{B^2 \exp(-B^2/2a^2)}{a^3} 
\end{eqnarray}
where 
\begin{eqnarray}
a = \frac{B_0}{\sqrt{3}}
\end{eqnarray}
and $B_0$ specifies the mean magnetic energy density, defined such that
\begin{eqnarray}
\int B^2 p_B dB = B_0^2
\end{eqnarray}

\subsection{Radio source evolution}  \label{sec:radio_source_evolution}

In the current work, we are focused on FRII radio galaxies, and therefore we assume that during the active phase the radio sources evolve according to the self-similar dynamical model of \citet{ka97}. During the remnant phase, the lobe evolution is not well understood, however, the true lobe expansion is likely to be bounded by two extreme cases of (i) maximal energy driven expansion and (ii) no expansion. For the maximal expansion rate of case (i) we consider the Sedov-like expansion described by \citet{kaiser02}. Sedov-like expansion refers to the situation in which the bow shock surrounding the lobes is driven by the adiabatic expansion of the lobes. This provides a solution that is similar to, but not entirely analogous to the Sedov solution for a point explosion, and is therefore referred to by Kaiser and Cotter as "Sedov-like". In order to apply the models of \citet{ka97} and \citet{kaiser02}, we must assume that the radio lobes expand into a power-law radial density profile, in which the ambient density, $\rho$, scales with the radial distance, $r$, according to
\begin{equation}
\rho \propto r^{-\beta}
\end{equation}

\subsubsection{Case (i), Sedov-like expansion in the remnant phase}

In our model, for case (i), the evolution of the radio source volume is a piece-wise power law, with:

\begin{equation}  \label{eqn:V_evolution}
  V(t) \propto\begin{cases}
    t^{9/(5 - \beta)}, & \text{if  } t<t_{\rm on}.\\
    t^{6/(6-\beta)}, & \text{if  } t>t_{\rm on}.
  \end{cases}
\end{equation}
where $t_{\rm on}$ is the length of the active phase. Our assumption of piece-wise power-law volume evolution means that the integrated electron energy distribution $N(\gamma)$ does not depend on the absolute value of the volume, only the exponents of the time evolution. For this reason, we have purposefully left the above expression as a proportionality. We assume that the magnetic field is isotropically distributed (``tangled") on all scales, and therefore can be treated as a magnetic ``fluid" with adiabatic index $\Gamma_B = 4/3$ \citep{leahy91}. In this case, the magnetic field evolves according to
\begin{equation}  \label{eqn:B_evolution}
  B(t)=\begin{cases}
    B_0 \left( \frac{t}{t_0} \right)^{\frac{-4-\beta}{2(5 - \beta)}} \left( \frac{Q}{Q_0} \right)^{\frac{2-\beta}{2(5 - \beta)}} , & \text{if  } t<t_{\rm on}.\\
    B(t_{\rm on}) \left( \frac{t}{t_{\rm on}} \right)^{\frac{-4}{(6-\beta)}}, & \text{if  } t>t_{\rm on}.
  \end{cases}
\end{equation}
\citep{ka97, kaiser02}, where $Q$ is the jet power and $Q_0 = 10^{39}$~W is a normalisation constant. We note that for $\beta \approx 1.5 - 2$, the dependence of magnetic field strength on jet power is extremely weak, and therefore $Q_0$ does not strongly affect the results. We assume that the particle energy distribution injected into the lobes is a power-law, such that 
\begin{equation} \label{eqn:injection_spectrum}
  \frac{dN}{d\gamma_i dt_i} =\begin{cases}
    q_0 \gamma_i^{-a} , & \text{if  } t<t_{\rm on} \text{  and  }  \gamma_{\rm i, min} < \gamma_i < \gamma_{\rm i, max}. \\
    0, & \text{otherwise  }
  \end{cases}
\end{equation}
where $q_0$ is proportional to the jet power, and a number of other assumed model parameters, as described in Section \ref{sec:sigma_and_jet_power}. 

\subsubsection{Case (ii), no expansion in the remnant phase}

The only difference in our model for case (i) and case (ii) is in equations \ref{eqn:V_evolution} and \ref{eqn:B_evolution} at $t>t_{\rm on}$. For completeness, they are defined below:
\begin{equation}  
  V(t) \propto\begin{cases}
    t^{9/(5 - \beta)}, & \text{if  } t<t_{\rm on}.\\
    \mbox{constant}, & \text{if  } t>t_{\rm on}.
  \end{cases}
\end{equation}
and
\begin{equation}  
  B(t)=\begin{cases}
    B_0 \left( \frac{t}{t_0} \right)^{\frac{-4-\beta}{2(5 - \beta)}} \left( \frac{Q}{Q_0} \right)^{\frac{2-\beta}{2(5 - \beta)}} , & \text{if  } t<t_{\rm on}.\\
    \mbox{constant}, & \text{if  } t>t_{\rm on}.
  \end{cases}
\end{equation}

\subsection{Volume integrated electron energy distribution}

Equations \ref{eqn:V_evolution}, \ref{eqn:B_evolution} and \ref{eqn:injection_spectrum} completely define our radio source model. The volume integrated electron energy distribution can be obtained from the completely general solution to the continuity equation (see Appendix \ref{app:vol_integrated_N_gamma}) 
\begin{equation}   \label{eqn:dN_dgamma}
\frac{dN}{d\gamma}\left( \gamma, t  \right)  = q_0 \gamma^{-a}   \int_{t_{\rm i, min}}^{t_{\rm i, max}}   \left(  \frac{V(t_i)}{V(t)} \right)^{(a-1)/3}  \left( 1 - \frac{\gamma}{\gamma_*(t_i, t)} \right)^{a-2} d t_i 
\end{equation}
where $t_{\rm i, max} = t$ for active sources, and $t_{\rm i, max} = t_{\rm on}$ for remnant sources. 
\begin{equation}   
\frac{1}{\gamma_*(t_i, t)} =  \int_{t_i}^{t}  a_0 \left(  \frac{V(\tau)}{V(t)}  \right)^{-1/3} \left( \frac{B^2(\tau) + B^2_{\rm CMB}}{2 \mu_0}  \right)  d\tau, 
\end{equation}
where $B_{\rm CMB} = 0.325 (1+z)^2$~nT is the equivalent magnetic field strength of the CMB, and the integration limit $t_{\rm i, min}$ is given by 
\begin{equation}  
t_{\rm i, min} = \mbox{MAX}(0, t_{\rm i, min}^*).
\end{equation}
The parameter $t_{\rm i, min}^*$ corresponds to the injection time at which a particle injected with Lorentz factor $\gamma_{\rm i, max}$ will have cooled to Lorentz factor $\gamma$ at time $t$, and is given by the solution to the equation
\begin{equation} 
\frac{1}{\gamma} = \frac{1}{\gamma_{\rm i, max} \left( \frac{V(t_{\rm i, min}^*)}{V(t)} \right)^{1/3}}   +   \frac{1}{\gamma_*(t_{\rm i, min}^*, t)} 
\end{equation}
where $\gamma_{\rm i, max}$ is the maximum electron Lorentz factor of the particle distribution injected into the lobes (see Appendix \ref{app:vol_integrated_N_gamma}). 

%We verified our procedure for calculating the volume integrated electron energy distribution by implementing a piece-wise linear finite volume method to solve the continuity equation numerically. 

\subsection{Simulation Approach} \label{sec:simulation_approach}

To generate a mock catalogue of radio galaxies, several of the model parameters are sampled from probability distributions, while others remain fixed. Each of the free parameters and the relevant distributions are discussed below. 

\subsubsection{Redshift}

Redshifts are sampled from a probability distribution given by 
\begin{equation}
p(z) \propto \rho(z) \frac{dV}{dz}
\end{equation}
where $\rho(z)$ is the volume density of radio galaxies as a function of redshift and $\frac{dV}{dz}$ is the differential comoving volume element for a spherical shell. For high power FRII radio galaxies, the volume density $\rho(z)$ is typically taken to be a Gaussian \citep{blundell99, willott01, grimes04}. We assume 
\begin{equation}
\rho(z) \propto \exp \left[ -\frac{1}{2}\left(  \frac{z - z_{h0}}{z_{h1}} \right) \right]
\end{equation}
with $z_{h0} = 1.95$ and $z_{h1} = 0.55$ \citep{grimes04}. The comoving volume element for a flat Universe ($\Omega_k = 0$) is 
\begin{equation}
\frac{dV}{dz} \propto \frac{(1+z)^2 D_A^2}{\sqrt{\Omega_M (1+z)^3 + \Omega_\Lambda}}
\end{equation}
\citep{hogg99}. 

\subsubsection{Radial density profile exponent, $\beta$}

As discussed, we assume that our mock radio galaxies expand into a power-law density profile with $\rho_{\rm ext} \propto r^{-\beta}$. We assume that the exponent $\beta = 1.9$ for all of our mock radio galaxies. Lower values of $\beta$ increase the fraction of remnants, but not by a large factor, due to the less rapid evolution of volume and magnetic field. 

\subsubsection{Active lifetime $t_{\rm on}$}

The typical active lifetime of FRII radio galaxies remains poorly constrained. Estimates based on self-similar dynamical models range from $\gtrsim$ 10 Myr \citep{bird08, kapinska12}, to 200 Myr \citep{antognini12}. 

Analysis of the length asymmetry of the most powerful double lobed radio sources indicates that the lobe advance speeds are typically a few percent of the speed of light, and not more than 0.15c \citep{scheuer95}, implying a typical active lifetime of a few tens of Myr.  

Estimates based on spectral ageing \citep{alexander87, liu92} systematically underestimate dynamical ages by a significant factor, and both dynamical and spectral age estimates have an uncertain relationship to the true source age \citep{eilek97, blundell00, kaiser05, hardcastle13a}. 

For the purposes of our simulated catalogues, we assume that the active lifetimes follow a truncated log-normal distribution, with mean $\langle \log(t_{\rm on}) \rangle = 7.5$ and standard deviation $\sigma_{\log(t_{\rm on})} = 0.1$, truncated such that $7.3 < \log(t_{\rm on}) < 8.3$, where $t_{\rm on}$ is specified in years. The active lifetime $t_{\rm on}$ refers to the time period during which the lobes are fed with fresh electrons. Increasing the mean active lifetime by a factor $f$ will cause a decrease in the simulated catalogue remnant fraction by a factor $\lesssim f$, while decreasing the mean active lifetime by a factor $f$ will increase the remnant fraction by a factor $\lesssim f$.

\subsubsection{Age at which the source is observed, $t_{\rm obs}$}

Source ages are sampled from a uniform distribution between $t_{\rm obs, min} = 0.1$ Myr and $t_{\rm obs, max} = 200$ Myr, which is several times the active lifetime. The results are insensitive to the assumed value for $t_{\rm obs, max}$, because the age distribution of remnants declines very steeply, as shown in Figure \ref{fig:histograms} (b). The results are insensitive to the assumed value for $t_{\rm obs, min}$, because $t_{\rm obs, min} \ll t_{\rm obs, max}$ and source age $t_{\rm obs}$ is sampled uniformly.  

\subsubsection{Magnetic field normalisation $B_0$}

One of the most important and influential parameters of our model is the normalisation of the magnetic field strength, $B_0$ in Equation \ref{eqn:B_evolution}. \citet{croston05} measured lobe magnetic field strengths in a sample of 33 powerful FRII radio galaxies, based on inverse Compton modelling of the observed X-ray emission. They find that the magnetic field strength in their sample is a strong function of source size. For sources greater than 300 kpc, \citet{croston05} find the median magnetic field strength is 0.6 nT, and reaches 0.2 nT for the largest sources in that sample. We therefore fix $B_0 = 0.6$ nT at time $\log(t_{\rm years}) = 7.3$ corresponding to the lower limit on the active lifetime for sources in our mock sample. In this way, we ensure that the magnetic field strength at the end of each source's active lifetime is $\lesssim 0.6$ nT, consistent with the results of \citet{croston05}. In equation \ref{eqn:B_evolution}, we have set $Q_0 = 10^{39}$ W.  However, we note that the dependence of magnetic field strength on jet power is extremely weak, and $Q_0$ is therefore not of great importance. 

\subsubsection{Energy injection index, $a$} \label{sec:injection_index_distribution}

The injection index is represented by the parameter $a$ in equation \ref{eqn:injection_spectrum}. We sample the injection index for each source from a truncated Gaussian distribution with $2.0 < a < 2.4$, mean $\bar{a} = 2.2$ and standard deviation $\sigma_a = 0.2$. This distribution of injection indices results in a relatively broad distribution of ``observed" spectral index in our mock catalogues, ranging between $0.5 < \alpha_{74}^{1400} < 1.2$ with a peak at $\alpha_{74}^{1400} \sim 0.8$, providing a good match to the spectral index distribution observed in our VLSSr sample. 

Note that with this injection index distribution, even the oldest active sources cannot have spectral index $\alpha > 1.2$ between any pair of frequencies, unless those frequencies are near to the frequency corresponding to the cutoff in the electron distribution. Therefore, active sources do not contaminate our ultra-steep spectrum sample in our mock catalogue. To explain this in more detail, consider a source with injection index $a=2.4$ -- the maximum possible value that could be drawn from the truncated Gaussian described above. Imagine this source is observed when it is very old, so that the radiative cooling break of $\Delta \alpha = 0.5$ occurs well below the lowest frequency we have observed. In that case, the spectral index will be $\alpha = 1.2$, but cannot get any steeper with age.

\subsubsection{Minimum/maximum injected Lorentz factor, $\gamma_{\rm min, max}$}

We assume $\gamma_{\rm min} = 100$ for all mock radio galaxies, and $\gamma_{\rm max} = 5 \times 10^6$. The maximum and minimum injected Lorentz factors do not strongly affect the results presented here, since the electrons at $\gamma_{\rm min}$ and those at $\gamma_{\rm max}$ emit at frequencies well outside the observed range. They do however affect the scaling between jet power and particle injection rate $q_0$ (see Section \ref{sec:sigma_and_jet_power}. ). 

Our choice for low energy electron cutoff is based on the observation in several sources of a low-frequency flattening in the hotspot spectra, consistent with a low energy cutoff in the electron distribution at a Lorentz factor of $\gamma_{\rm min} \sim 100 - 700$ \citep[][and references therein]{godfrey09}. Such values of $\gamma_{\rm min}$ are likely to be the result of dissipation of jet bulk kinetic energy \citep{godfrey09}. In the case of Cygnus A, absorption is at least partially responsible for the low frequency turnover in the hotspot spectra \citep{mckean16}, however, this does not rule out the possible involvement of a low energy cutoff in the hotspots of Cygnus A. The interpretation of the hotspot turnover in Cygnus A remains problematic \citep{mckean16}.

\subsubsection{Jet power, $Q_{\rm jet}$}

The jet power is sampled from a power-law probability distribution, with
\begin{equation}
p(Q_{\rm jet}) \propto \begin{cases} 
Q^{-n_Q} \text{ if } 5 \times 10^{36} ~ W < Q_{\rm jet} < 2 \times 10^{42} ~ W \\
= 0 \text{  otherwise}
\end{cases}
\end{equation}
We assume $n_Q = 2.3$, which is an average of the values derived by \citet{blundell99} (2.6) and \citet{wang08} (2.0). We note that the results presented here are not strongly dependent on the assumed value of $n_Q$, for reasonable departures from our assumed value.

\subsubsection{Electron energy fraction, $\epsilon_e$}

To calculate the particle injection rate $q_0$, we must specify the fraction of jet power that is converted to the internal energy of the relativistic electron population, $\epsilon_e$ (see Equation \ref{eqn:q_0}). Here we assume that the jet power is equally distributed between magnetic energy, relativistic electron energy, and the energy of non-radiating particles, so that $\epsilon_e = 1/3$. Our conclusions are not sensitive to the value of $\epsilon_e$. 

\subsubsection{Ratio of hotspot pressure to lobe pressure, $\frac{p_{\rm HS}}{p_{\rm lobe}}$}

Another parameter that is necessary to calculate the particle injection rate into the lobes, is the ratio of hotspot pressure to lobe pressure (see Equation \ref{eqn:q_0}). Here we assume $\frac{p_{\rm HS}}{p_{\rm lobe}} = 10$, however we note the very weak dependence of $q_0$ on the value of $\frac{p_{\rm HS}}{p_{\rm lobe}}$. Again, our conclusions are not sensitive to the assumed value of $\frac{p_{\rm HS}}{p_{\rm lobe}}$.

\subsection{Contribution of the jets and hotspots in active sources}

After the jets stop supplying energy to the hotspots, the bright emission from them will disappear in the order of a sound-crossing time ($\lesssim 1$~Myr for a hotspot with a diameter of a few kpc). If the hotspots provide a significant fraction of the total source flux at 74 MHz, then the rapid disappearance of the hotspots at the start of the remnant phase could help to explain the rapid disappearance of remnant FRII radio galaxies from our flux limited samples. 

\citet{jenkins77} defined the hotspots as regions 15 kpc in diameter, and found that the ratio of hotspot to total luminosity is strongly correlated with the total luminosity, and furthermore, that the hotspots can often contribute more than 90$\%$ of the source flux density at 178 MHz, particularly at high luminosities. 

In direct contrast to this result, is the more recent study by \citet{mullin08} of a sample of 100 FRII 3CRR radio galaxies with $z < 1$. \citet{mullin08} are able to more accurately identify the hotspot regions than \citet{jenkins77}, and find no correlation between hotspot prominence and the total radio luminosity in this sample (see their figure 32). Furthermore, \citet{mullin08} find that the hotspots contribute typically a few percent to a few tens of percent to the total radio luminosity at 178 MHz\footnote{\citet{mullin08} define hotspot prominence as the ratio of hotspot luminosity at 8.4 GHz to the total source luminosity at 178MHz. We have converted the hotspot prominence of \citet{mullin08} to a hotspot emission fraction, or compactness in the terminology of \citet{jenkins77} by multiplying the hotspot prominence values by a factor of $(8400/178)^{-\alpha}$ assuming $\alpha = 0.8$.}. 

The median hotspot prominence (summed for both the north and south hotspot) for the sample of \citet{mullin08} is 0.008. We convert this to a hotspot fraction (the ratio of total source luminosity contributed by the hotspots at 74 MHz) by multiplying the hotspot prominence by a factor of $(74/8400)^{-\alpha}$ assuming $\alpha = 0.8$. This gives a median hotspot fraction of 35$\%$. However, this does not account for the fact that hotspot spectra often become much flatter towards low frequency, particularly in high luminosity radio galaxies \citep{leahy89, godfrey09, mckean16}. We therefore do not attempt to account for the contribution to the source flux from hotspot related emission. We simply note that if the source flux density drops by a factor $f_{\rm drop}$ at the end of the active phase due to the rapid disappearance of the hotspots, then we will over-estimate the fraction of remnant to active sources by a factor of $f_{\rm drop}^{p-1}$, where $p \approx 2.3$ is the slope of the luminosity function.

The jet prominence is typically much lower than the hotspot prominence, by an order of magnitude or more \citep{mullin08}. While it is true that in a few percent of cases, the jet prominence is comparable or greater than the hotspot prominence, in the vast majority of cases, the jet prominence is a negligible component of the total source flux density.

\subsection{Procedure for generating mock catalogues}

The mock catalogues are created in several steps, as described below. 

\begin{enumerate}
\item First, we generate several million radio galaxies, each with the model parameters fixed or sampled from their corresponding probability distributions as described in the preceding sections.
\item For each source, we calculate an upper limit to the flux density at the sample selection frequency (74 MHz) using a fast, analytic expression.
\item We apply the flux cut to the sample, using the upper limits calculated in the previous step: All sources for which the calculated flux upper limit at the selection frequency is below the flux limit are removed from the mock catalogue. 
\item For the remaining sources, we calculate the model radio galaxy spectrum accurately using numerical integration of equations \ref{eqn:dN_dgamma} and \ref{eqn:S_nu}. 
\item We again apply a flux cut to the sample, this time using the accurate model flux densities as the basis of the flux cut. Only those sources whose flux densities lie above the flux limit remain in the sample. 
\end{enumerate}

 The flux density upper limit in step (ii) described above, is obtained by using an analytic approximation to equation \ref{eqn:dN_dgamma}, along with the $\delta-$function approximation to the synchrotron emission spectrum. The analytic approximation to equation \ref{eqn:dN_dgamma} is derived by neglecting the term $(1 - \gamma/\gamma^*)^{a-2}$. That is, we replace the following equation (Equation \ref{eqn:dN_dgamma})
\begin{equation} 
\frac{dN}{d\gamma}\left( \gamma, t  \right)  = q_0 \gamma^{-a}   \int_{t_{\rm i, min}}^{t_{\rm i, max}}   \left(  \frac{V(t_i)}{V(t)} \right)^{(a-1)/3}  \left( 1 - \frac{\gamma}{\gamma_*(t_i, t)} \right)^{a-2} d t_i  \nonumber
\end{equation}
with the following integral, 
\begin{equation}  \label{eqn:upper_limit_on_N_gamma}
\frac{dN}{d\gamma}\left( \gamma, t  \right)  < q_0 \gamma^{-a}   \int_{t_{\rm i, min}}^{t_{\rm i, max}}   \left(  \frac{V(t_i)}{V(t)} \right)^{(a-1)/3}  d t_i  
\end{equation}
which in the case of power law volume evolution, has an analytic solution. Since the term $(1 - \gamma/\gamma^*)^{a-2} < 1$, Equation \ref{eqn:upper_limit_on_N_gamma} provides an upper limit on $N(\gamma)$, and therefore an upper limit on the flux density at any frequency.

\section{Modeling Results}  \label{sec:modeling_results}

We carried out two Monte-Carlo simulations, each one representing a bound on the possible evolution scenarios in the remnant phase, and each one differing only in the prescribed evolution of lobe volume and magnetic field strength in the remnant phase (see the description of these scenarios in section \ref{sec:radio_source_evolution}). 
In the remnant phase, for case (i), we assume that the lobes remain over pressured with respect to the ambient medium, and evolve in a Sedov-like manner as described by \citet{kaiser02}. This is the maximal expansion rate that can be achieved by inactive radio galaxies. In case (ii) we assume that there is no lobe expansion in the remnant phase. 

In the active phase, for both of our simulations, we assume the lobes evolve according to self-similar expansion models of \citet{ka97}. 

We note that our model is only applicable to radio galaxies with a single active phase: re-started radio galaxies will show different characteristic luminosity and spectral evolution.

\subsection{Results of Monte-Carlo simulation with Maximal (sedov-like) expansion in the remnant phase (Case (i))}

\begin{table*}
 \caption{Results of Monte Carlo Simulations: All redshifts.}
 \label{table:modeling_results_all_redshifts}
  \begin{tabular}{cccccc}
  \hline
  & \multicolumn{2}{c}{Maximal (Sedov-like) expansion} && \multicolumn{2}{c}{No expansion} \\
  & \multicolumn{2}{c}{in remnant phase} && \multicolumn{2}{c}{in remnant phase} \\
  \cline{2-3} \cline{5-6} \\
  Quantity & Number of sources & Fraction of sample && Number of sources & Fraction of sample  \\
  \hline
Active sources & 9499 & 94.6$\%$ && 5372 & 82$\%$ \\
\\
All Remnant sources & 539 & 5.4$\%$ && 1199 & 18$\%$  \\
\\
Low $\nu$ ultra-steep spectrum remnants ($\alpha_{\rm 74~MHz}^{\rm 1400 MHz} > 1.2$) & 187 & 1.8$\%$  && 751 & 11.4$\%$ \\
\\
Mid $\nu$ ultra-steep spectrum remnants ($\alpha_{\rm 1400 MHz}^{\rm 5 GHz} > 1.2$) & 378 & 3.8$\%$  && 1036 & 15.8$\%$ \\
\\
High $\nu$ ultra-steep spectrum remnants ($\alpha_{\rm 5~GHz}^{\rm 10 GHz} > 1.2$) & 441 & 4.4$\%$  && 1110 & 16.9$\%$ \\
\\
Curved spectrum remnants ($\alpha_{\rm 1400~MHz}^{\rm 5000~MHz} - \alpha_{\rm 74~MHz}^{\rm 1400~MHz} > 0.5$) & 292 & 2.9$\%$ && 976 & 14.9$\%$ \\
\\
Curved spectrum remnants ($\alpha_{\rm 5000~MHz}^{\rm 10000~MHz} - \alpha_{\rm 74~MHz}^{\rm 1400~MHz} > 0.5$) & 385 & 3.8$\%$ && 1068 & 16.3$\%$ \\
  \hline
\end{tabular}
\end{table*}

\begin{table*}
 \caption{Results of Monte Carlo Simulations: Low redshifts only ($z<0.5$).}
 \label{table:modeling_results_low_redshifts}
  \begin{tabular}{cccccc}
  \hline
  & \multicolumn{2}{c}{Maximal (Sedov-like) expansion} && \multicolumn{2}{c}{No expansion} \\
  & \multicolumn{2}{c}{in remnant phase} && \multicolumn{2}{c}{in remnant phase} \\
  \cline{2-3} \cline{5-6} \\
  Quantity & Number of sources & Fraction of sample && Number of sources & Fraction of sample  \\
  \hline
Active sources & 496 & 80$\%$   && 292 & 37$\%$ \\
\\
All Remnant sources & 123 & 20$\%$   && 505 & 63$\%$\\
\\
Low $\nu$ ultra-steep spectrum remnants ($\alpha_{\rm 74~MHz}^{\rm 1400 MHz} > 1.2$) & 20 & 3.2$\%$  && 265 & 33$\%$ \\
\\
Mid $\nu$ ultra-steep spectrum remnants ($\alpha_{\rm 1400~MHz}^{\rm 5 GHz} > 1.2$) & 57 & 9.2$\%$  && 419 & 53$\%$ \\
\\
High $\nu$ ultra-steep spectrum remnants ($\alpha_{\rm 5~GHz}^{\rm 10 GHz} > 1.2$) & 79 & 12.8$\%$  && 456 & 57$\%$ \\
\\
Curved spectrum remnants ($\alpha_{\rm 1400~MHz}^{\rm 5000~MHz} - \alpha_{\rm 74~MHz}^{\rm 1400~MHz} > 0.5$) & 47 & 7.6$\%$ && 397 & 50$\%$ \\
\\
Curved spectrum remnants ($\alpha_{\rm 5000~MHz}^{\rm 10000~MHz} - \alpha_{\rm 74~MHz}^{\rm 1400~MHz} > 0.5$) & 69 & 11.1$\%$ && 448 & 56$\%$  \\
  \hline
\end{tabular}
\end{table*}

\begin{figure*}
%	\includegraphics[width=2.0\columnwidth]{figures/2D_panels_plot_327MHz.pdf}
   	\includegraphics[width=2.0\columnwidth]{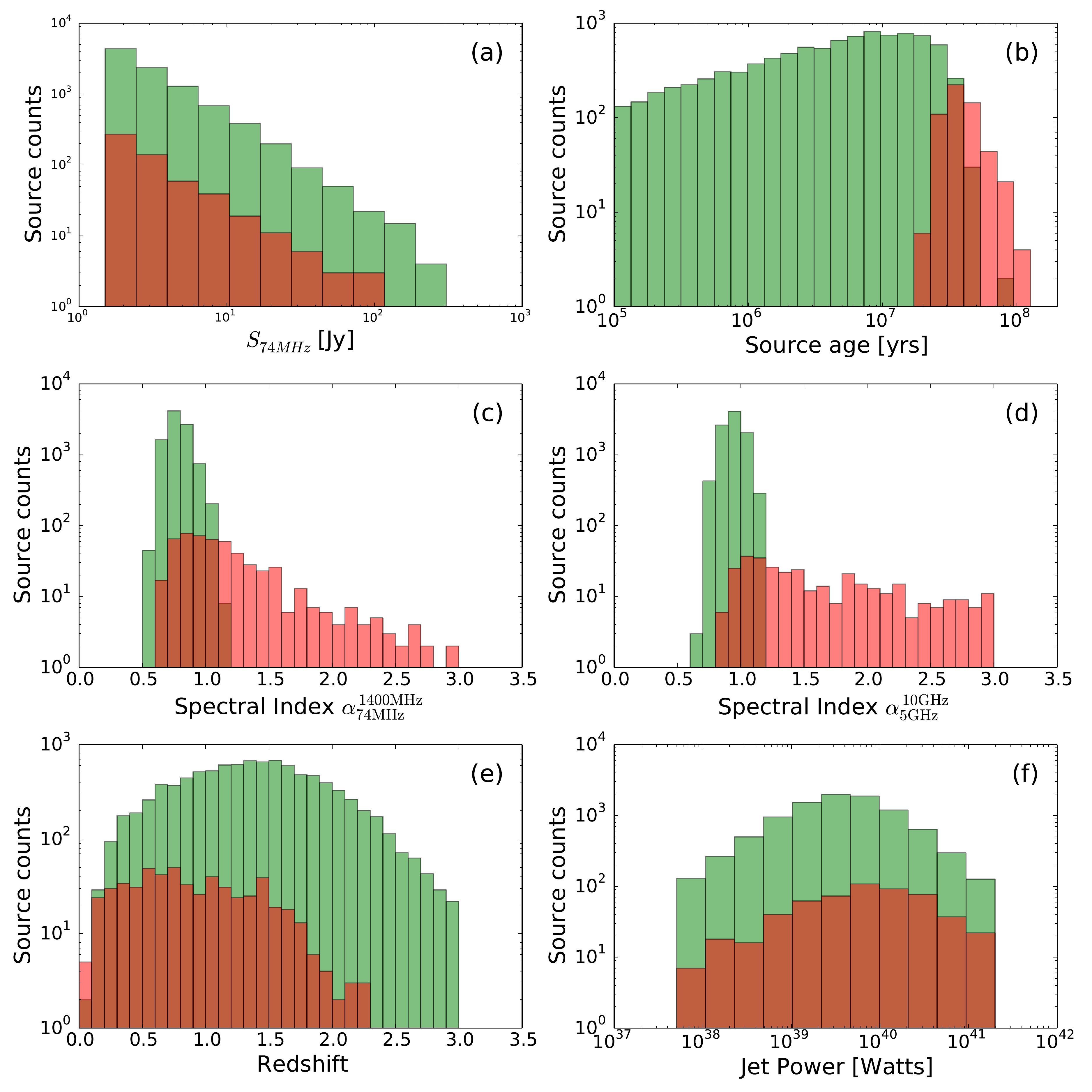}
    \caption{Histograms of spectral index and flux densities for our mock catalogue obtained under the assumption of Sedov-like expansion in the remnant phase (case (i)). Green bars represent active sources, red bars represent remnant sources. Note that not all remnant sources are ultra-steep spectrum (see Figure \ref{fig:remnant_fraction_vs_redshift}).}
    \label{fig:histograms}
 %-----
 % This plot produced by /Users/godfrey/astro/Projects/Current_Focus/1_Remnant_RGs_and_Generalised_CI_model/VLSSr_x_NVSS_Sample_Selection/Get_VLSSxNVSS_sample.py 
 %-----
 \end{figure*}

\begin{figure}
%	\includegraphics[width=2.0\columnwidth]{figures/2D_panels_plot_327MHz.pdf}
   	\includegraphics[width=1.0\columnwidth]{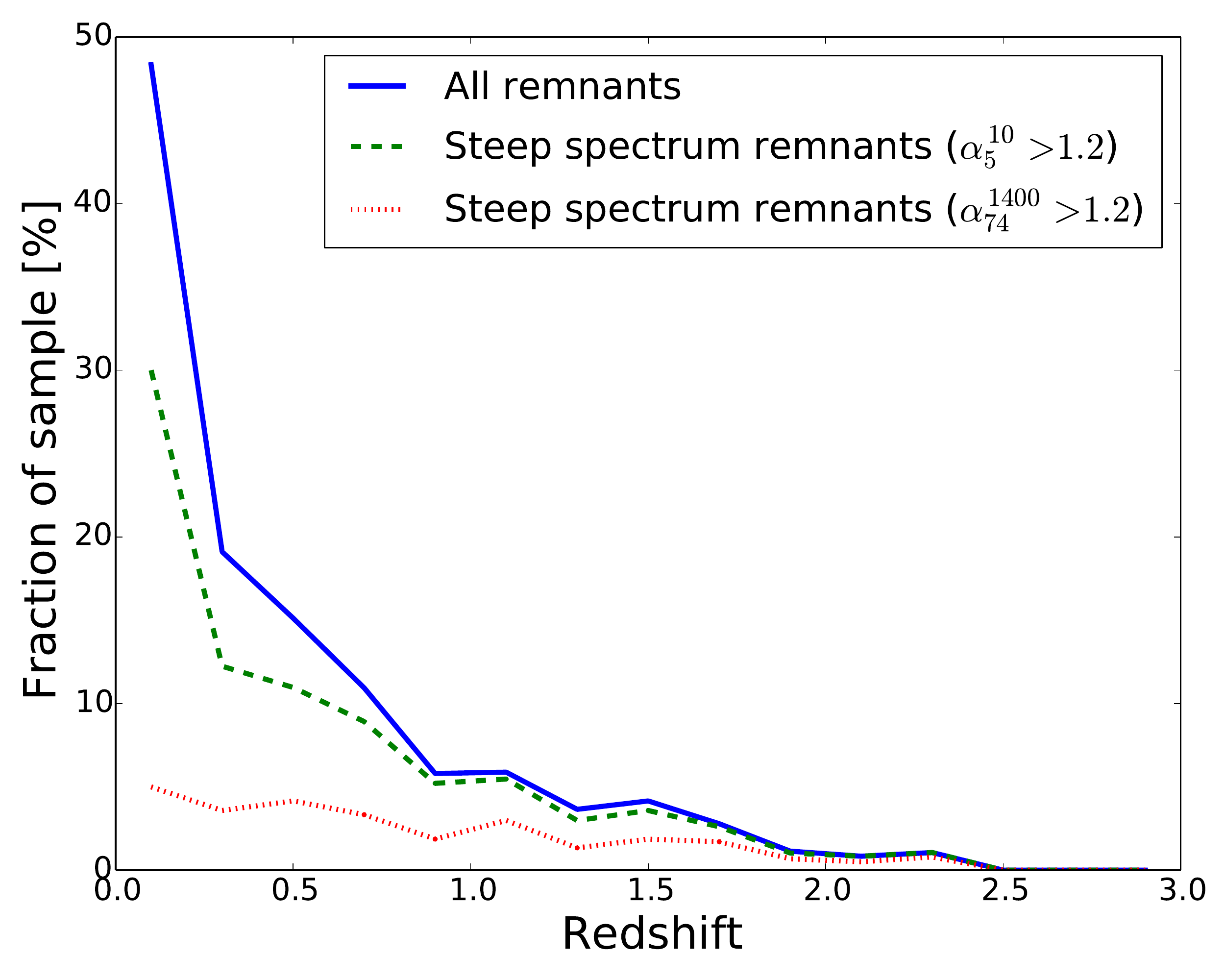}
    \caption{Fraction of flux limited sample that are remnants, as a function of redshift, for our simulated population of $S_{\rm 74 MHz} > 1.5$~Jy FRII radio galaxies, assuming Sedov-like expansion in the remnant phase.}
    \label{fig:remnant_fraction_vs_redshift}
 %-----
 % This plot produced by /Users/godfrey/astro/Projects/Current_Focus/1_Remnant_RGs_and_Generalised_CI_model/VLSSr_x_NVSS_Sample_Selection/Get_VLSSxNVSS_sample.py 
 %-----
 \end{figure}
In Figures \ref{fig:histograms} and \ref{fig:remnant_fraction_vs_redshift} and Tables \ref{table:modeling_results_all_redshifts} and \ref{table:modeling_results_low_redshifts} we summarise the results of our simulations of active and remnant radio galaxy populations with maximal (sedov-like) expansion in the remnant phase.

Our simulated catalogue contains a total of 10,038 sources\footnote{We purposefully simulated a larger catalogue of sources than was obtained in our VLSSr selected sample in order to increase the fidelity of the simulated distributions.}, of which 9499 are active (94.6$\%$), and 539 are remnant (5.4$\%$). However, of the 539 remnants, only 187 (1.8$\%$ of the entire sample) have ultra-steep spectra between 74 MHz and 1400 MHz with $\alpha_{\rm 74 MHz}^{\rm 1400 MHz} > 1.2$. The fraction of ultra-steep spectrum remnants in our mock sample is therefore consistent with our upper limit of $2\%$ on the ultra-steep spectrum FRII remnant fraction obtained in section \ref{sec:empirical_results}. However, this model predicts almost two times as many remnants would have been missed by our remnant selection criterion. The 74 MHz to 1400 MHz spectral index is clearly an inefficient selection criterion, with a selection efficiency of only $35\%$ (187/539). Spectral indices measured at higher frequency are significantly more efficient at remnant selection. If remnant selection incorporated higher frequency data using the criterion $\alpha_{\rm 1.4 GHz}^{\rm 5 GHz} > 1.2$, our Monte-Carlo simulation predicts that we would more than double the number of remnants selected, and achieve $70\%$ remnant selection efficiency. If we could have incorporated both 5GHz and 10 GHz data into our analysis, and selected remnants based on the criterion $\alpha_{\rm 5 GHz}^{\rm 10 GHz} > 1.2$, our Monte-Carlo simulation predicts a selection efficiency of 82$\%$.

One of the problems with selecting remnants based on ultra-steep spectra alone is that there are other types of radio source that can satisfy the ultra-steep spectrum criterion, such as high-redshift radio galaxies (HzRGs). To overcome this problem, some authors have used spectral curvature as a more robust indicator of remnant radio galaxies \citep[eg.][]{murgia11}. Spectral curvature is calculated as the difference between a high frequency spectral index and low frequency spectral index, and requires flux density measurements at at least three different frequencies. We have considered the spectral curvature selection in our mock sample using measurements at three frequencies, up to 5 GHz ($\alpha_{\rm 1400~MHz}^{\rm 5000~MHz} - \alpha_{\rm 74~MHz}^{\rm 1400~MHz} > 0.5$), and also using four frequencies up to 10 GHz ($\alpha_{\rm 1400~MHz}^{\rm 5000~MHz} - \alpha_{\rm 74~MHz}^{\rm 1400~MHz} > 0.5$), achieving remnant selection efficiency of 54$\%$ and $71\%$, respectively.

Figure \ref{fig:histograms} (a) demonstrates that the flux density distribution for the remnant population has the same slope as that of the active population, despite their different luminosity evolution. 
This implies that the overall remnant fraction is independent of the flux limit of the sample. 

Figure \ref{fig:histograms} (b) demonstrates the very sharp decline in the number of remnants as a function of source age. This implies that remnants in flux limited samples tend to be relatively ``new" remnants, as expected due to the rapid luminosity evolution that arises from a combination of decreasing magnetic field strength along with the adiabatic and radiative losses during the remnant phase. This tendency for remnant radio galaxies in flux limited samples to be young is the reason that many remnants in our Monte-Carlo simulations do not have ultra-steep spectra in the observed frequency range. Most importantly, unlike the luminosity distribution, the age distribution of a flux limited sample of remnants can provide meaningful constraints on the luminosity evolution during the remnant phase. To illustrate this, we consider the following simplified example. Suppose that sources are ``born" with some peak luminosity $L_{\rm peak}$, and evolve according to a piecewise power law with 
\begin{eqnarray} \label{eqn:L_obs}
L_\nu &\propto& L_{\rm peak} t^{-a_1} \qquad \qquad \qquad t < t_{\rm on}  \nonumber \\
L_\nu &\propto& L_{\rm peak} t^{-a_2} \quad \qquad ~ t > t_{\rm on}  \nonumber
\end{eqnarray}
Now assume that sources are ``born" with luminosity $L_{\rm peak}$ at a rate given by the birth function
\begin{eqnarray}  \label{eqn:L_peak_pdf}
\frac{dN}{dt ~ d L_{\rm peak}} &\propto&  L_{\rm peak}^{-p}  \nonumber
\end{eqnarray}
Then the age distribution of remnant radio galaxies at a given luminosity $L_\nu$ is given by 
\begin{equation}
\frac{dN}{dt d L_\nu} \propto t^{a_2 (1 - p)}
\end{equation}
Thus, given an estimate of the ``birth" luminosity function (i.e. $p$ in the above model), the age distribution can be used to constrain the luminosity evolution in the remnant phase.

Figures \ref{fig:histograms} (c) and (d) demonstrate the existence of many remnant radio galaxies with ``normal" spectral index values in our mock catalogue, as well as the broad tail of ultra-steep spectrum remnants.

Figure \ref{fig:histograms} (e) demonstrates that high redshift remnant radio galaxies are extremely rare, and low redshift remnant radio galaxies are predicted to be much more common. This is due to the combination of two factors: (1) the increased rest-frame frequency at higher redshifts corresponds to higher energy electrons, and consequently corresponds to a faster radiative cooling rate; and (2) the increased energy density of the CMB at higher redshifts causes a faster radiative cooling rate from inverse Compton scattering of the CMB. The number of remnants per redshift bin is approximately constant up to $z \gtrsim 1$, but decreases dramatically for $z \gtrsim 1.5$. In Table \ref{table:modeling_results_low_redshifts} we list results for the low redshift ($z<0.5$) subsample of our simulated catalogues. In the simulated catalogue with Sedov-like expansion in the remnant phase, 20$\%$ of the flux limited FRII sample is predicted to be remnant, and of those remnants, approximately half will show ultra-steep spectra between 1.4 GHz and 5 GHz. 
It is clear from Table \ref{table:modeling_results_low_redshifts} and Figures \ref{fig:histograms} (e) and \ref{fig:remnant_fraction_vs_redshift} that there is a significant advantage to be gained by studying the remnant population at low redshifts, particularly when high frequencies (5 and 10 GHz) are included in the analysis.

\subsection{Results of Monte-Carlo simulation with no expansion in the remnant phase}

We repeated the Monte-Carlo simulations with no expansion in the remnant phase, so that 
\begin{equation}  \label{eqn:V_evolution_no_sedov}
  V(t) \propto\begin{cases}
    t^{9/(5 - \beta)}, & \text{if  } t<t_{\rm on}.\\
    V(t_{\rm on}), & \text{if  } t>t_{\rm on}.
  \end{cases}
\end{equation}
and
\begin{equation}  \label{eqn:B_evolution_no_sedov}
  B(t)=\begin{cases}
    B_0 \left( \frac{t}{t_0} \right)^{\frac{-4-\beta}{2(5 - \beta)}} \left( \frac{Q}{Q_0} \right)^{\frac{2-\beta}{2(5 - \beta)}} , & \text{if  } t<t_{\rm on}.\\
    B(t_{\rm on}), & \text{if  } t>t_{\rm on}.
  \end{cases}
\end{equation}
This model is not physically realistic: it is unlikely that all sources will reach pressure equilibrium right at the moment the central engine shuts off. However, it is a useful exercise to demonstrate the strength of the effect of expansion in the remnant phase. It also clearly demonstrates that models with no expansion in the remnant phase are in great conflict with observations. This is important, given the evidence that FRII radio galaxies might reach pressure equilibrium with the surroundings before the end of their active life. 

\begin{figure}
%	\includegraphics[width=2.0\columnwidth]{figures/2D_panels_plot_327MHz.pdf}
   	\includegraphics[width=1.0\columnwidth]{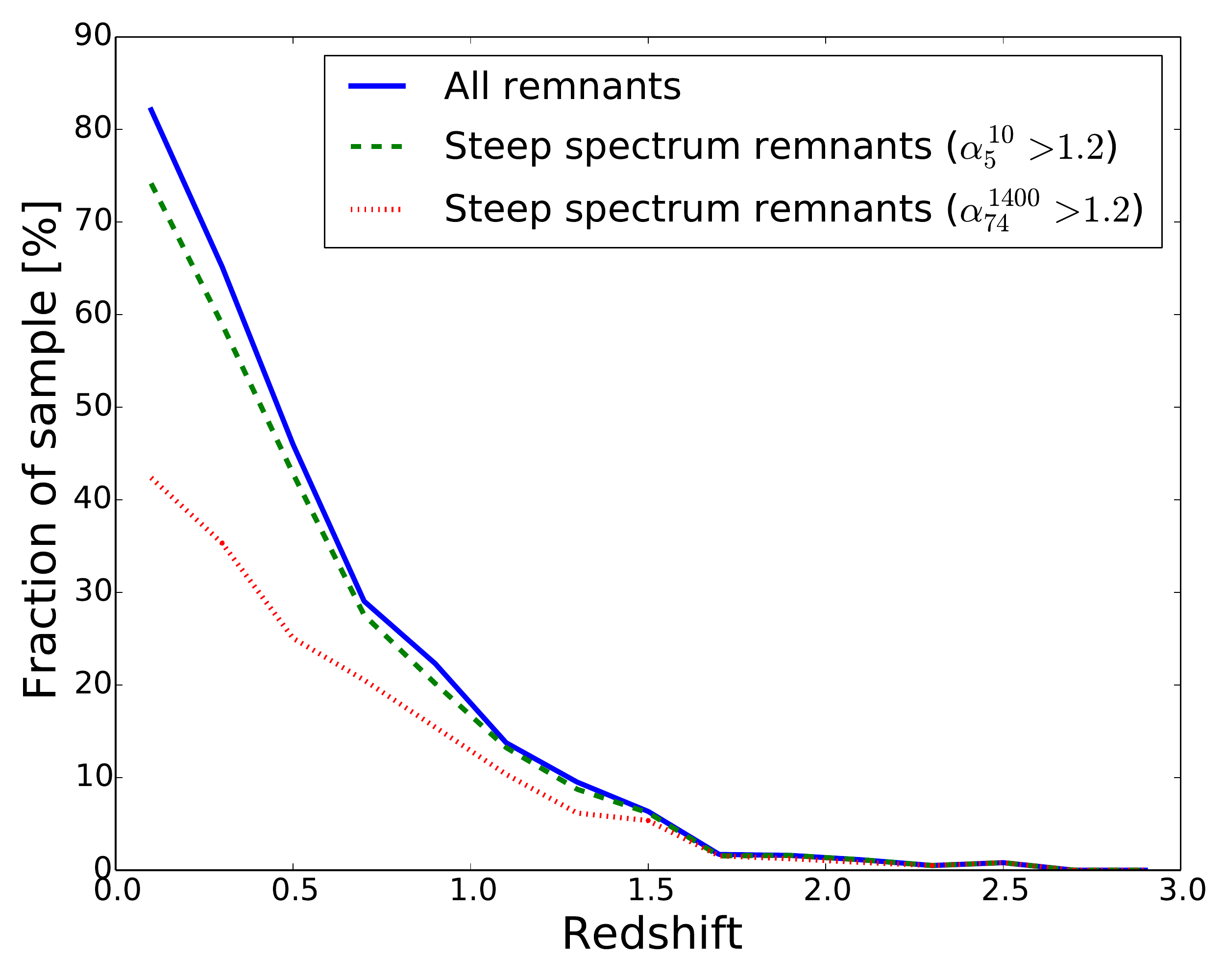}
    \caption{Fraction of flux limited sample that are remnants, as a function of redshift, for our simulated population of $S_{\rm 74 MHz} > 1.5$~Jy FRII radio galaxies, assuming there are no adiabatic losses, and no magnetic field evolution in the remnant phase.}
    \label{fig:remnant_fraction_vs_redshift_no_expansion_in_remnant_phase}
 %-----
 % This plot produced by /Users/godfrey/astro/Projects/Current_Focus/1_Remnant_RGs_and_Generalised_CI_model/VLSSr_x_NVSS_Sample_Selection/Get_VLSSxNVSS_sample.py 
 %-----
 \end{figure}

The results for this model with no expansion in the remnant phase are presented in figures \ref{fig:histograms_no_expansion_in_remnant_phase} and \ref{fig:remnant_fraction_vs_redshift_no_expansion_in_remnant_phase} and Tables \ref{table:modeling_results_all_redshifts} and \ref{table:modeling_results_low_redshifts}. It is immediately clear, as expected, that the remnant fraction is greater in this simulated catalogue, particularly at low redshift ($z < 0.5$), where the model predicts that nearly two-thirds of the flux limited sample are remnants, and that a third of the flux limited sample are ultra-steep spectrum remnants with $\alpha_{\rm 74~MHz}^{\rm 1400~MHz} > 1.2$ (see table \ref{table:modeling_results_low_redshifts} and Figure \ref{fig:remnant_fraction_vs_redshift_no_expansion_in_remnant_phase}).

\begin{figure}
%	\includegraphics[width=2.0\columnwidth]{figures/2D_panels_plot_327MHz.pdf}
   	\includegraphics[width=1.0\columnwidth]{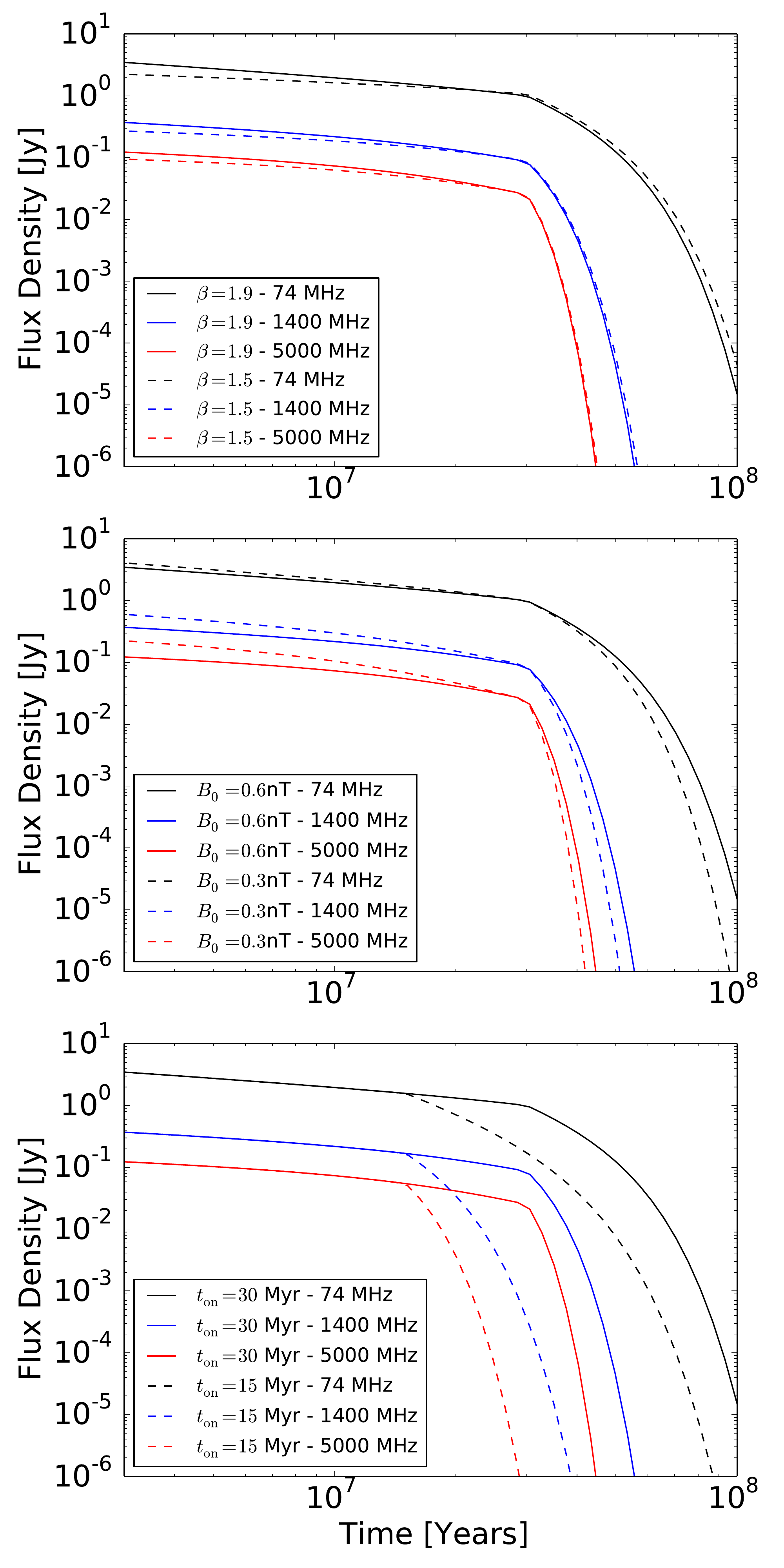}
    \caption{This figure illustrates the effect of three different model parameters on the flux density evolution of model radio galaxies in the active and remnant phase, at 74~MHz, 1.4~GHz, and 5~GHz. The three model parameters of interest in this figure are the most significant in terms of their effect on the remnant fraction derived from our Monte Carlo simulations. In each figure the model parameters are fixed except for either $\beta$, $B_0$ or $t_{\rm on}$, the values of which are specified in the figure legend. The model parameters are as follows, unless specified otherwise in the figure legend: 
$\gamma_{\rm min} = 100$, 
$\gamma_{\rm max} = 5 \times 10^6$, 
$\epsilon_e = 1/3$, 
$\frac{p_{\rm hs}}{p_{\rm lobe}} = 10$, 
$B_0 = 0.6$ nT, 
$Q_0 = 10^{39}$, 
$z = 1.0$, 
$t_{\rm on} = 30$ Myr, 
$\beta = 1.9$, 
$Q_{\rm jet} = 10^{39}$ Watts, 
injection index $a = 2.2$. 
See section \ref{sec:simulation_approach} for a description of each of these model parameters. 
Note that in the second panel, each of the model curves for $B_0 = 0.3$nT have been scaled by a factor $\gtrsim$ 3, to enable better comparison between the light curves.} 
    \label{fig:luminosity_evolution_comparison}
 %-----
 % This plot produced by /Users/godfrey/astro/Projects/Current_Focus/1_Remnant_RGs_and_Generalised_CI_model/AAA_Mock_Catalogues/create_luminosity_evolution_plots_for_paper/luminosity_evolution_for_paper_make_plots.py
 %-----
 \end{figure}

\begin{figure*}
%	\includegraphics[width=2.0\columnwidth]{figures/2D_panels_plot_327MHz.pdf}
   	\includegraphics[width=2.0\columnwidth]{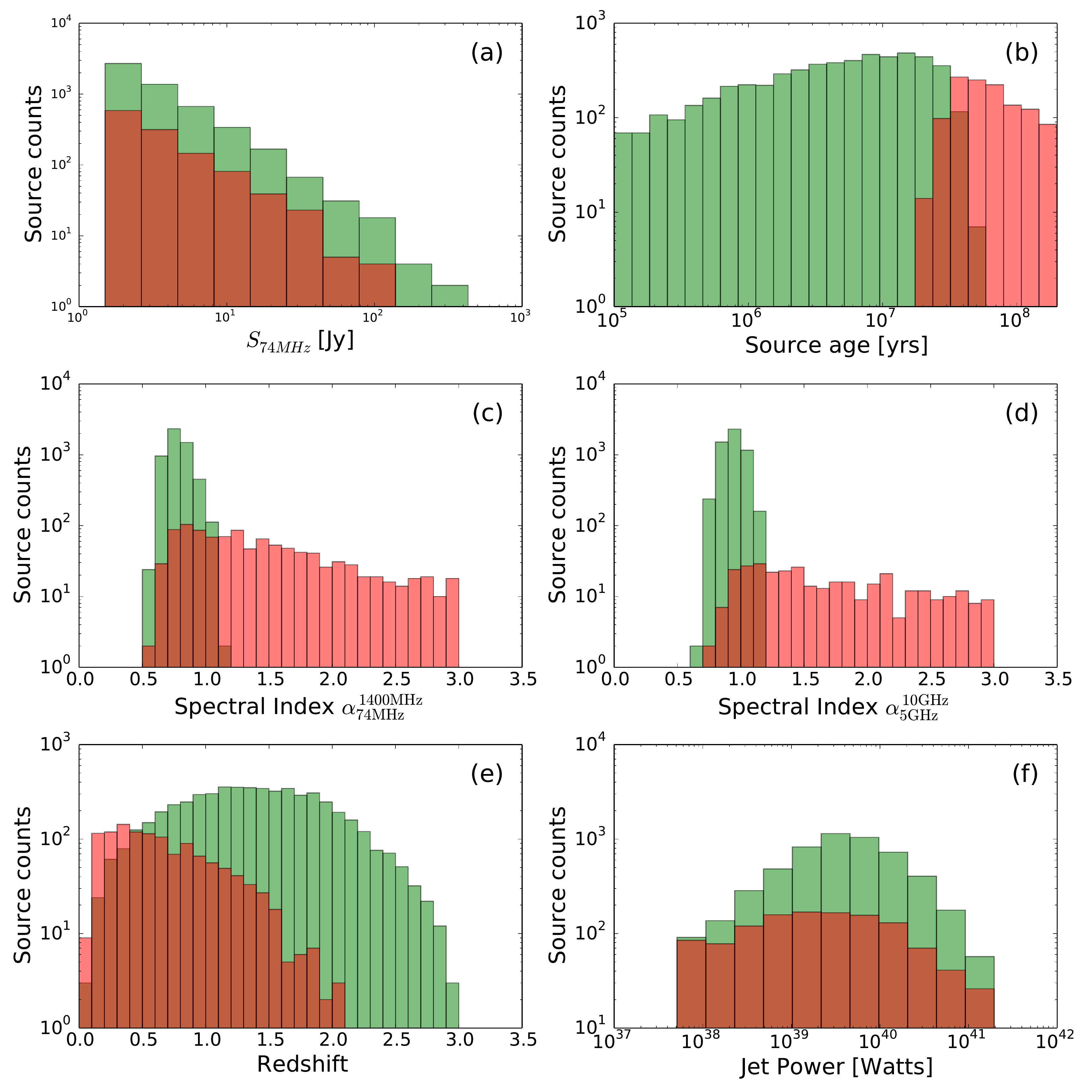}
    \caption{Histograms of spectral index and flux densities for our mock catalogue under the assumption of no expansion in the remnant phase (case (ii)). Green bars represent active sources, red bars represent remnant sources. Note that not all remnant sources are ultra-steep spectrum (see Figure \ref{fig:remnant_fraction_vs_redshift}).}
    \label{fig:histograms_no_expansion_in_remnant_phase}
 %-----
 % This plot produced by /Users/godfrey/astro/Projects/Current_Focus/1_Remnant_RGs_and_Generalised_CI_model/VLSSr_x_NVSS_Sample_Selection/Get_VLSSxNVSS_sample.py 
 %-----
 \end{figure*}

\newpage

\section{Discussion and Conclusions}

We have carried out an empirical study based on a flux limited sample from the VLSSr radio catalogue, in order to place a limit on the occurrence of remnant FRII radio galaxies in a 74 MHz flux limited sample. We have also performed Monte-Carlo simulations of the population of active and remnant FRII radio galaxies to assess whether models of remnant lobe evolution are consistent with the observed remnant fraction. Our main conclusions may be summarised as follows:

\begin{enumerate}
\item In our VLSSr selected sample, fewer than 2$\%$ of FRII radio galaxies with 74 MHz flux density greater than 1.5 Jy are ultra-steep spectrum remnants with  $\alpha_{\rm 74~MHz}^{\rm 1400~MHz} > 1.2$. 
\item Our Monte-Carlo simulation with Maximal (sedov-like) expansion in the remnant phase produced a remnant fraction of $2\%$, marginally consistent with the upper limit described above. 
\item Our Monte-Carlo simulation with Maximal (sedov-like) expansion in the remnant phase predicts the existence of nearly twice as many remnants with ``normal" spectra ($\alpha_{\rm 74~MHz}^{\rm 1400~MHz} < 1.2$) as there are ultra-steep spectrum remnants in our sample. 
\item The above conclusion can be phrased another way: the ultra-steep selection criterion $\alpha_{\rm 74~MHz}^{\rm 1400~MHz} > 1.2$ is not efficient at selecting remnant radio galaxies, with a selection efficiency of only $\sim 35\%$. Higher frequency spectral indices are significantly more efficient at remnant selection. Remnant selection based on the criterion $\alpha_{\rm 1.4 GHz}^{\rm 5 GHz} > 1.2$ increases the selection efficiency to $70\%$, and remnant selection based on the criterion $\alpha_{\rm 5 GHz}^{\rm 10 GHz} > 1.2$ increases the selection efficiency to 82$\%$. For redshifts less than 0.5, the number of identified ultra-steep spectrum remnants increases by a factor of 4 when using $\alpha_{\rm 5~GHz}^{\rm 10~GHz}$ as opposed to $\alpha_{\rm 74~MHz}^{\rm 1400~MHz}$ (Table \ref{table:modeling_results_low_redshifts}). 
\item The remnant fraction increases rapidly towards low redshift. This is the result of (1) the increased rest-frame frequency at higher redshifts and (2) the increased energy density of the CMB at higher redshifts, resulting in and increase in the radiative cooling rate from inverse Compton scattering of the CMB. 
\item The model predicts an ultra-steep remnant fraction approaching 10$\%$ at redshifts $z < 0.5$, when considering ultra-steep selection based on higher frequency data ($\alpha_{\rm 1.4~GHz}^{\rm 5~GHz} > 1.2$).
\item The age distribution of remnant radio galaxies in flux limited samples is a steeply decreasing function of sources age, indicating that most remnant radio galaxies in flux limited samples are young remnants (Figures \ref{fig:histograms} (b) and     \ref{fig:histograms_no_expansion_in_remnant_phase} (b)). Due to the steep remnant age distribution, incorporating higher frequency data into the analysis will be more important in future studies than incorporating lower frequencies. 
\item The age distribution of remnant radio galaxies in a flux limited sample can constrain the luminosity evolution of the remnant phase. The luminosity distribution of remnants cannot. 
\item The spectral index distribution of remnant radio galaxies peaks at ``normal" spectral index values. This is a direct result of the age distribution: most remnants are young in flux limited samples (Figure \ref{fig:histograms} (c)). 
\item In the idealised situation that we have modelled, and with the selection criteria we have used to identify candidate remnants, the high frequency spectral index is more efficient at selecting remnants than the spectral curvature. However, when considering heterogeneous samples as obtained for example from flux limited samples, the spectral curvature is likely to be more robust and may result in fewer false remnant candidates. 
\item The remnant fraction is independent of flux limit (Figure \ref{fig:histograms} (a)). Therefore, going to fainter flux limits will not increase the remnant fraction. 
\end{enumerate}

As discussed in Section \ref{sec:introduction}, several studies suggest that FRII radio lobes may reach pressure equilibrium before the end of their lifetime. However, we have shown that models without rapid remnant phase expansion significantly over-predict the FRII remnant fraction. Rapid luminosity evolution in the remnant phase resulting from Sedov-like expansion is required to match the low observed remnant fraction in our flux limited sample. Our results imply that either the previous evidence for internal/external pressure equilibrium in FRII radio galaxy lobes is flawed, or alternative mechanisms other than adiabatic expansion are required explain the low remnant fraction in our flux limited sample.

\section*{Acknowledgements}

The research leading to these results has received funding from the European Research Council under the European Union's Seventh Framework Programme (FP/2007-2013) / ERC Advanced Grant RADIOLIFE-320745.

\bibliographystyle{mnras}
\bibliography{LEHG_Refs} % if your bibtex file is called LEHG_Refs.bib

\begin{thebibliography}{}
\makeatletter
\relax
\def\mn@urlcharsother{\let\do\@makeother \do\$\do\&\do\#\do\^\do\_\do\%\do\~}
\def\mn@doi{\begingroup\mn@urlcharsother \@ifnextchar [ {\mn@doi@}
  {\mn@doi@[]}}
\def\mn@doi@[#1]#2{\def\@tempa{#1}\ifx\@tempa\@empty \href
  {http://dx.doi.org/#2} {doi:#2}\else \href {http://dx.doi.org/#2} {#1}\fi
  \endgroup}
\def\mn@eprint#1#2{\mn@eprint@#1:#2::\@nil}
\def\mn@eprint@arXiv#1{\href {http://arxiv.org/abs/#1} {{\tt arXiv:#1}}}
\def\mn@eprint@dblp#1{\href {http://dblp.uni-trier.de/rec/bibtex/#1.xml}
  {dblp:#1}}
\def\mn@eprint@#1:#2:#3:#4\@nil{\def\@tempa {#1}\def\@tempb {#2}\def\@tempc
  {#3}\ifx \@tempc \@empty \let \@tempc \@tempb \let \@tempb \@tempa \fi \ifx
  \@tempb \@empty \def\@tempb {arXiv}\fi \@ifundefined
  {mn@eprint@\@tempb}{\@tempb:\@tempc}{\expandafter \expandafter \csname
  mn@eprint@\@tempb\endcsname \expandafter{\@tempc}}}

\bibitem[\protect\citeauthoryear{{Abramowitz} \& {Stegun}}{{Abramowitz} \&
  {Stegun}}{1970}]{abramowitz70}
{Abramowitz} M.,  {Stegun} I.~A.,  1970, {Handbook of mathematical functions :
  with formulas, graphs, and mathematical tables}

\bibitem[\protect\citeauthoryear{{Alexander} \& {Leahy}}{{Alexander} \&
  {Leahy}}{1987}]{alexander87}
{Alexander} P.,  {Leahy} J.~P.,  1987, \mn@doi [\mnras]
  {10.1093/mnras/225.1.1}, \href
  {http://adsabs.harvard.edu/abs/1987MNRAS.225....1A} {225, 1}

\bibitem[\protect\citeauthoryear{{Antognini}, {Bird}  \& {Martini}}{{Antognini}
  et~al.}{2012}]{antognini12}
{Antognini} J.,  {Bird} J.,   {Martini} P.,  2012, \mn@doi [\apj]
  {10.1088/0004-637X/756/2/116}, \href
  {http://adsabs.harvard.edu/abs/2012ApJ...756..116A} {756, 116}

\bibitem[\protect\citeauthoryear{{Bird}, {Martini}  \& {Kaiser}}{{Bird}
  et~al.}{2008}]{bird08}
{Bird} J.,  {Martini} P.,   {Kaiser} C.,  2008, \mn@doi [\apj]
  {10.1086/527534}, \href {http://adsabs.harvard.edu/abs/2008ApJ...676..147B}
  {676, 147}

\bibitem[\protect\citeauthoryear{{Blundell} \& {Rawlings}}{{Blundell} \&
  {Rawlings}}{2000}]{blundell00}
{Blundell} K.~M.,  {Rawlings} S.,  2000, \mn@doi [\aj] {10.1086/301254}, \href
  {http://adsabs.harvard.edu/abs/2000AJ....119.1111B} {119, 1111}

\bibitem[\protect\citeauthoryear{Blundell, Rawlings  \& Willott}{Blundell
  et~al.}{1999}]{blundell99}
Blundell K.~M.,  Rawlings S.,   Willott C.~J.,  1999, The Astronomical Journal,
  117, 677

\bibitem[\protect\citeauthoryear{{Brienza} et~al.,}{{Brienza}
  et~al.}{2016}]{brienza16}
{Brienza} M.,  et~al., 2016, \mn@doi [\aap] {10.1051/0004-6361/201526754},
  \href {http://adsabs.harvard.edu/abs/2016A%26A...585A..29B} {585, A29}

\bibitem[\protect\citeauthoryear{{Cohen}, {Lane}, {Cotton}, {Kassim}, {Lazio},
  {Perley}, {Condon}  \& {Erickson}}{{Cohen} et~al.}{2007}]{cohen07}
{Cohen} A.~S.,  {Lane} W.~M.,  {Cotton} W.~D.,  {Kassim} N.~E.,  {Lazio}
  T.~J.~W.,  {Perley} R.~A.,  {Condon} J.~J.,   {Erickson} W.~C.,  2007,
  \mn@doi [\aj] {10.1086/520719}, \href
  {http://adsabs.harvard.edu/abs/2007AJ....134.1245C} {134, 1245}

\bibitem[\protect\citeauthoryear{{Condon}}{{Condon}}{1997}]{condon97}
{Condon} J.~J.,  1997, \mn@doi [\pasp] {10.1086/133871}, \href
  {http://adsabs.harvard.edu/abs/1997PASP..109..166C} {109, 166}

\bibitem[\protect\citeauthoryear{{Condon}, {Cotton}, {Greisen}, {Yin},
  {Perley}, {Taylor}  \& {Broderick}}{{Condon} et~al.}{1998}]{condon98}
{Condon} J.~J.,  {Cotton} W.~D.,  {Greisen} E.~W.,  {Yin} Q.~F.,  {Perley}
  R.~A.,  {Taylor} G.~B.,   {Broderick} J.~J.,  1998, \mn@doi [\aj]
  {10.1086/300337}, \href {http://adsabs.harvard.edu/abs/1998AJ....115.1693C}
  {115, 1693}

\bibitem[\protect\citeauthoryear{{Croston}, {Birkinshaw}, {Hardcastle}  \&
  {Worrall}}{{Croston} et~al.}{2004}]{croston04}
{Croston} J.~H.,  {Birkinshaw} M.,  {Hardcastle} M.~J.,   {Worrall} D.~M.,
  2004, \mn@doi [\mnras] {10.1111/j.1365-2966.2004.08118.x}, \href
  {http://adsabs.harvard.edu/abs/2004MNRAS.353..879C} {353, 879}

\bibitem[\protect\citeauthoryear{{Croston}, {Hardcastle}, {Harris}, {Belsole},
  {Birkinshaw}  \& {Worrall}}{{Croston} et~al.}{2005}]{croston05}
{Croston} J.~H.,  {Hardcastle} M.~J.,  {Harris} D.~E.,  {Belsole} E.,
  {Birkinshaw} M.,   {Worrall} D.~M.,  2005, \mn@doi [\apj] {10.1086/430170},
  \href {http://adsabs.harvard.edu/abs/2005ApJ...626..733C} {626, 733}

\bibitem[\protect\citeauthoryear{Crusius \& Schlickeiser}{Crusius \&
  Schlickeiser}{1986}]{crusius86}
Crusius A.,  Schlickeiser R.,  1986, Astronomy and Astrophysics (ISSN
  0004-6361), 164, L16

\bibitem[\protect\citeauthoryear{{Eilek}, {Melrose}  \& {Walker}}{{Eilek}
  et~al.}{1997}]{eilek97}
{Eilek} J.~A.,  {Melrose} D.~B.,   {Walker} M.~A.,  1997, \apj, \href
  {http://adsabs.harvard.edu/abs/1997ApJ...483..282E} {483, 282}

\bibitem[\protect\citeauthoryear{En{\ss}lin \& Gopal-Krishna}{En{\ss}lin \&
  Gopal-Krishna}{2001}]{ensslin01}
En{\ss}lin T.~A.,  Gopal-Krishna 2001, A\&A, 366, 26

\bibitem[\protect\citeauthoryear{{En{\ss}lin}, {Lieu}  \&
  {Biermann}}{{En{\ss}lin} et~al.}{1999}]{ensslin99}
{En{\ss}lin} T.~A.,  {Lieu} R.,   {Biermann} P.~L.,  1999, \aap, \href
  {http://adsabs.harvard.edu/abs/1999A%26A...344..409E} {344, 409}

\bibitem[\protect\citeauthoryear{Godfrey et~al.,}{Godfrey
  et~al.}{2009}]{godfrey09}
Godfrey L. E.~H.,  et~al., 2009, 695, 707

\bibitem[\protect\citeauthoryear{{Grimes}, {Rawlings}  \& {Willott}}{{Grimes}
  et~al.}{2004}]{grimes04}
{Grimes} J.~A.,  {Rawlings} S.,   {Willott} C.~J.,  2004, \mn@doi [\mnras]
  {10.1111/j.1365-2966.2004.07510.x}, \href
  {http://adsabs.harvard.edu/abs/2004MNRAS.349..503G} {349, 503}

\bibitem[\protect\citeauthoryear{Hardcastle}{Hardcastle}{2013}]{hardcastle13a}
Hardcastle M.~J.,  2013, \mnras, 433, 3364

\bibitem[\protect\citeauthoryear{{Hardcastle} \& {Worrall}}{{Hardcastle} \&
  {Worrall}}{2000}]{hardcastle00}
{Hardcastle} M.~J.,  {Worrall} D.~M.,  2000, \mn@doi [\mnras]
  {10.1046/j.1365-8711.2000.03883.x}, \href
  {http://adsabs.harvard.edu/abs/2000MNRAS.319..562H} {319, 562}

\bibitem[\protect\citeauthoryear{{Hardcastle}, {Birkinshaw}, {Cameron},
  {Harris}, {Looney}  \& {Worrall}}{{Hardcastle} et~al.}{2002}]{hardcastle02}
{Hardcastle} M.~J.,  {Birkinshaw} M.,  {Cameron} R.~A.,  {Harris} D.~E.,
  {Looney} L.~W.,   {Worrall} D.~M.,  2002, \mn@doi [\apj] {10.1086/344409},
  \href {http://adsabs.harvard.edu/abs/2002ApJ...581..948H} {581, 948}

\bibitem[\protect\citeauthoryear{{Hogg}}{{Hogg}}{1999}]{hogg99}
{Hogg} D.~W.,  1999, astro-ph/9905116, \href
  {http://adsabs.harvard.edu/abs/1999astro.ph..5116H} {}

\bibitem[\protect\citeauthoryear{{Jenkins} \& {McEllin}}{{Jenkins} \&
  {McEllin}}{1977}]{jenkins77}
{Jenkins} C.~J.,  {McEllin} M.,  1977, \mn@doi [\mnras]
  {10.1093/mnras/180.2.219}, \href
  {http://adsabs.harvard.edu/abs/1977MNRAS.180..219J} {180, 219}

\bibitem[\protect\citeauthoryear{{Kaiser}}{{Kaiser}}{2005}]{kaiser05}
{Kaiser} C.~R.,  2005, \mn@doi [\mnras] {10.1111/j.1365-2966.2005.09022.x},
  \href {http://adsabs.harvard.edu/abs/2005MNRAS.360..176K} {360, 176}

\bibitem[\protect\citeauthoryear{Kaiser \& Alexander}{Kaiser \&
  Alexander}{1997}]{ka97}
Kaiser C.~R.,  Alexander P.,  1997, \mnras, 286, 215

\bibitem[\protect\citeauthoryear{{Kaiser} \& {Cotter}}{{Kaiser} \&
  {Cotter}}{2002}]{kaiser02}
{Kaiser} C.~R.,  {Cotter} G.,  2002, \mn@doi [\mnras]
  {10.1046/j.1365-8711.2002.05799.x}, \href
  {http://adsabs.harvard.edu/abs/2002MNRAS.336..649K} {336, 649}

\bibitem[\protect\citeauthoryear{Kaiser, Dennett-Thorpe  \& Alexander}{Kaiser
  et~al.}{1997}]{kda97}
Kaiser C.~R.,  Dennett-Thorpe J.,   Alexander P.,  1997, \mnras, 292, 723

\bibitem[\protect\citeauthoryear{Kapi{\'{n}}ska, Uttley  \&
  Kaiser}{Kapi{\'{n}}ska et~al.}{2012}]{kapinska12}
Kapi{\'{n}}ska A.~D.,  Uttley P.,   Kaiser C.~R.,  2012, \mnras, p.~3301

\bibitem[\protect\citeauthoryear{Komissarov \& Gubanov}{Komissarov \&
  Gubanov}{1994}]{komissarov94}
Komissarov S.~S.,  Gubanov A.~G.,  1994, A\&A, 285, 27

\bibitem[\protect\citeauthoryear{{Lane}, {Cotton}, {van Velzen}, {Clarke},
  {Kassim}, {Helmboldt}, {Lazio}  \& {Cohen}}{{Lane} et~al.}{2014}]{lane14}
{Lane} W.~M.,  {Cotton} W.~D.,  {van Velzen} S.,  {Clarke} T.~E.,  {Kassim}
  N.~E.,  {Helmboldt} J.~F.,  {Lazio} T.~J.~W.,   {Cohen} A.~S.,  2014, \mn@doi
  [\mnras] {10.1093/mnras/stu256}, \href
  {http://adsabs.harvard.edu/abs/2014MNRAS.440..327L} {440, 327}

\bibitem[\protect\citeauthoryear{{Leahy}}{{Leahy}}{1991}]{leahy91}
{Leahy} J.~P.,  1991, {Interpretation of large scale extragalactic jets}.
p.~100

\bibitem[\protect\citeauthoryear{{Leahy}, {Muxlow}  \& {Stephens}}{{Leahy}
  et~al.}{1989}]{leahy89}
{Leahy} J.~P.,  {Muxlow} T.~W.~B.,   {Stephens} P.~W.,  1989, \mn@doi [\mnras]
  {10.1093/mnras/239.2.401}, \href
  {http://adsabs.harvard.edu/abs/1989MNRAS.239..401L} {239, 401}

\bibitem[\protect\citeauthoryear{{Liu}, {Pooley}  \& {Riley}}{{Liu}
  et~al.}{1992}]{liu92}
{Liu} R.,  {Pooley} G.,   {Riley} J.~M.,  1992, \mn@doi [\mnras]
  {10.1093/mnras/257.4.545}, \href
  {http://adsabs.harvard.edu/abs/1992MNRAS.257..545L} {257, 545}

\bibitem[\protect\citeauthoryear{{Manolakou} \& {Kirk}}{{Manolakou} \&
  {Kirk}}{2002}]{manolakou02}
{Manolakou} K.,  {Kirk} J.~G.,  2002, \mn@doi [\aap]
  {10.1051/0004-6361:20020780}, \href
  {http://adsabs.harvard.edu/abs/2002A%26A...391..127M} {391, 127}

\bibitem[\protect\citeauthoryear{{Massardi}, {Bonaldi}, {Negrello},
  {Ricciardi}, {Raccanelli}  \& {de Zotti}}{{Massardi}
  et~al.}{2010}]{massardi10}
{Massardi} M.,  {Bonaldi} A.,  {Negrello} M.,  {Ricciardi} S.,  {Raccanelli}
  A.,   {de Zotti} G.,  2010, \mn@doi [\mnras]
  {10.1111/j.1365-2966.2010.16305.x}, \href
  {http://adsabs.harvard.edu/abs/2010MNRAS.404..532M} {404, 532}

\bibitem[\protect\citeauthoryear{{McKean} et~al.,}{{McKean}
  et~al.}{2016}]{mckean16}
{McKean} J.~P.,  et~al., 2016, \mn@doi [\mnras] {10.1093/mnras/stw2105}, \href
  {http://adsabs.harvard.edu/abs/2016MNRAS.463.3143M} {463, 3143}

\bibitem[\protect\citeauthoryear{Mullin, Riley  \& Hardcastle}{Mullin
  et~al.}{2008}]{mullin08}
Mullin L.~M.,  Riley J.~M.,   Hardcastle M.~J.,  2008, \mnras, 390, 595

\bibitem[\protect\citeauthoryear{Murgia, Fanti, Fanti, Gregorini, Klein, Mack
  \& Vigotti}{Murgia et~al.}{1999}]{murgia99}
Murgia M.,  Fanti C.,  Fanti R.,  Gregorini L.,  Klein U.,  Mack K.~H.,
  Vigotti M.,  1999, Astronomy and Astrophysics, 345, 769

\bibitem[\protect\citeauthoryear{{Murgia} et~al.,}{{Murgia}
  et~al.}{2011}]{murgia11}
{Murgia} M.,  et~al., 2011, \mn@doi [\aap] {10.1051/0004-6361/201015302}, \href
  {http://adsabs.harvard.edu/abs/2011A%26A...526A.148M} {526, A148}

\bibitem[\protect\citeauthoryear{{Orienti}, {Prieto}, {Brunetti}, {Mack},
  {Massaro}  \& {Harris}}{{Orienti} et~al.}{2012}]{orienti12}
{Orienti} M.,  {Prieto} M.~A.,  {Brunetti} G.,  {Mack} K.-H.,  {Massaro} F.,
  {Harris} D.~E.,  2012, \mn@doi [\mnras] {10.1111/j.1365-2966.2011.19882.x},
  \href {http://adsabs.harvard.edu/abs/2012MNRAS.419.2338O} {419, 2338}

\bibitem[\protect\citeauthoryear{{Scheuer}}{{Scheuer}}{1995}]{scheuer95}
{Scheuer} P.~A.~G.,  1995, \mn@doi [\mnras] {10.1093/mnras/277.1.331}, \href
  {http://adsabs.harvard.edu/abs/1995MNRAS.277..331S} {277, 331}

\bibitem[\protect\citeauthoryear{{Tribble}}{{Tribble}}{1991}]{tribble91}
{Tribble} P.~C.,  1991, \mn@doi [\mnras] {10.1093/mnras/253.1.147}, \href
  {http://adsabs.harvard.edu/abs/1991MNRAS.253..147T} {253, 147}

\bibitem[\protect\citeauthoryear{{Tribble}}{{Tribble}}{1993}]{tribble93}
{Tribble} P.~C.,  1993, \mn@doi [\mnras] {10.1093/mnras/261.1.57}, \href
  {http://adsabs.harvard.edu/abs/1993MNRAS.261...57T} {261, 57}

\bibitem[\protect\citeauthoryear{{Wang} \& {Kaiser}}{{Wang} \&
  {Kaiser}}{2008}]{wang08}
{Wang} Y.,  {Kaiser} C.~R.,  2008, \mn@doi [\mnras]
  {10.1111/j.1365-2966.2008.13417.x}, \href
  {http://adsabs.harvard.edu/abs/2008MNRAS.388..677W} {388, 677}

\bibitem[\protect\citeauthoryear{Willott, Rawlings, Blundell, Lacy  \&
  Eales}{Willott et~al.}{2001}]{willott01}
Willott C.~J.,  Rawlings S.,  Blundell K.~M.,  Lacy M.,   Eales S.~A.,  2001,
  \mnras, 322, 536

\bibitem[\protect\citeauthoryear{{Wilman} et~al.,}{{Wilman}
  et~al.}{2008}]{wilman08}
{Wilman} R.~J.,  et~al., 2008, \mn@doi [\mnras]
  {10.1111/j.1365-2966.2008.13486.x}, \href
  {http://adsabs.harvard.edu/abs/2008MNRAS.388.1335W} {388, 1335}

\bibitem[\protect\citeauthoryear{van Velzen, Falcke  \& K{\"o}rding}{van Velzen
  et~al.}{2015}]{vanvelzen15}
van Velzen S.,  Falcke H.,   K{\"o}rding E.,  2015, \mnras, 446, 2985

\makeatother
\end{thebibliography}

%%%%%%%%%%%%%%%%%%%%%%%%%%%%%%%%%%%%%%%%

%%%%%%%%%%%%%%%%% APPENDICES %%%%%%%%%%%%%%%%%%%%%

\newpage

\appendix

\section{The Generalised Continuous Injection Model for Lobed Radio Galaxies}  \label{}

We describe a mathematical framework for modelling the integrated spectra of radio galaxies in various phases of their life, under the following set of assumptions. 

\begin{enumerate}
\item The electron energy distribution is independent of the local magnetic energy density. 
\item The cooling rate is a function of time and particle energy only. Specifically, this assumption implies that the cooling rate is independent of position within the lobes. This is a common assumption used in modelling integrated radio spectra, and is valid provided that the average magnetic energy density (averaged over the cooling length scale) is independent of position within the lobes. 
\end{enumerate}

This mathematical framework allows for arbitrary evolution of lobe volume and magnetic field strength, it allows for arbitrary, and time-dependent injection spectra (important for modelling objects in which jet power evolves with time), and it allows for a distribution of magnetic field strengths within each volume element of the lobes.

However, this type of model is only valid for lobed radio galaxies, in which the emission is dominated by the lobes, and the mean magnetic energy density is approximately constant throughout the lobe volume. This model is therefore not suited to the modelling of tailed or naked jet FRI radio galaxies. 

We present a simplified version of the generalised model, which is relevant to modelling objects with multiple power-law phases of lobe evolution, such as remnant radio galaxies, as well as cluster radio-relics \citep{ensslin01}. 

\subsection{Integrated flux density}

Consider a small volume of lobe plasma $\Delta V$ in which the number density of relativistic particles per unit volume and per unit Lorentz factor is given by $\frac{dN}{d\gamma dV} = n(\gamma)$, and in which the magnetic field is described by $p_\Theta$ and $p_B$, where $p_\Theta$ and $p_B$ are the probability distributions of the pitch angle $\Theta$ and the magnetic field strength $B$, respectively. \citet{hardcastle13a} give an expression for the volume-averaged emissivity from such a volume element \citep[see][for details]{hardcastle13a}:

\begin{equation}  \label{eqn:j_nu_appendix_eqn}
j_\nu = \int_0^\infty \int_0^\pi \int_{\gamma_{\rm min}}^{\gamma_{\rm max}} \frac{\sqrt{3} B e^3 \sin \Theta}{8 \pi^2 \epsilon_0 c m_e} F(x) n(\gamma) p_\Theta p_B d\gamma d\Theta dB 
\end{equation}
where $e$ and $m_e$ are the charge and mass of the electron, respectively, $c$ is the speed of light in vacuum, $\epsilon_0$ is the permttivity of free space, $x = \frac{\nu}{\nu_c}$, $\nu$ is the rest-frame frequency, $\nu_c = \frac{3}{4 \pi} \Omega_0 \gamma^2 \sin \Theta$ is the characteristic synchrotron frequency, $\Omega_0$ is the non-relativistic gyrofrequency which in SI units is given by $\Omega_0 = e B/m_e$, and the synchrotron function $F(x) = x \int_x^{\infty} K_{5/3}(z)dz$, where $K_{5/3}(z)$ is the modified Bessel function of order 5/3.

If we assume that $n(\gamma)$, $B$ and $\Theta$ are all independent (assumption (i)), then we can separate the integrals, and the equation becomes
\begin{equation}
j_\nu = \frac{\sqrt{3} e^3 }{8 \pi^2 \epsilon_0 c m_e} \int_{\gamma_{\rm min}}^{\gamma_{\rm max}} n(\gamma)  \left[ \int_0^\infty   B p_B \left[ \int_0^\pi \sin \Theta F(x)  p_\Theta  d\Theta \right] dB \right] d\gamma
\end{equation}
Let us define the parameter 
\begin{equation}
y = \frac{\nu \sin \Theta}{\nu_c} = \frac{4 \pi \nu}{3 \Omega_0 \gamma^2}.
\end{equation}
We can write the inner integral over pitch-angle as (see Appendix \ref{app:F_bar_y})
\begin{equation}
\bar{F}(y) = \int_0^\pi \sin \Theta F(x)  p_\Theta  d\Theta. 
\end{equation}
Then
\begin{equation}
j_\nu = \frac{\sqrt{3} e^3 }{8 \pi^2 \epsilon_0 c m_e} \int_{\gamma_{\rm min}}^{\gamma_{\rm max}} n(\gamma)  \left[ \int_0^\infty   B p_B \bar{F} (y) dB \right] d\gamma
\end{equation}

The flux density of the lobe is 
\begin{equation}
S_\nu = \frac{(1+z)}{D_L^2} \int j^\prime_{\nu^\prime} dV^\prime
\end{equation}
\begin{equation}
= \frac{(1+z)}{D_L^2} \frac{\sqrt{3} e^3 }{8 \pi^2 \epsilon_0 c m_e} \int_{V^\prime}  \int_{\gamma_{\rm min}}^{\gamma_{\rm max}} n(\gamma)  \left[ \int_0^\infty   B p_B \bar{F} (y) dB \right] d\gamma  dV^\prime 
\end{equation}

If we assume that the magnetic field distribution $p_B$ and $p_\Theta$ is the same within each volume element $\Delta V$ throughout the lobes (assumption (ii)),  then we can switch the order of integration, so that
\begin{equation}
S_\nu = \frac{(1+z)}{D_L^2} \frac{\sqrt{3} e^3 }{8 \pi^2 \epsilon_0 c m_e} \int_{\gamma_{\rm min}}^{\gamma_{\rm max}} \int_{V^\prime} n(\gamma)dV^\prime   \left[ \int_0^\infty   B p_B \bar{F} (y) dB \right] d\gamma 
\end{equation}
Let us define
\begin{equation}
N(\gamma) = \frac{dN}{d\gamma} = \int n(\gamma) dV
\end{equation}
Then
\begin{equation}
S_\nu = \frac{(1+z)}{D_L^2} \frac{\sqrt{3} e^3 }{8 \pi^2 \epsilon_0 c m_e} \int_{\gamma_{\rm min}}^{\gamma_{\rm max}} N(\gamma) \left[ \int_0^\infty   B p_B \bar{F} (y) dB \right]d\gamma
\end{equation}

For a Guassian-random field, the probability distribution for the magnetic field strength $p_B$ is the Maxwell-Boltzmann distribution \citep[][]{hardcastle13a}:
\begin{eqnarray}
p_B = \sqrt{\frac{2}{\pi}} \frac{B^2 \exp(-B^2/2a^2)}{a^3} 
\end{eqnarray}
where 
\begin{eqnarray}
a = \frac{B_0}{\sqrt{3}}
\end{eqnarray}
and $B_0$ specifies the mean magnetic energy density, and is defined such that
\begin{eqnarray}
\int B^2 p_B dB = B_0^2
\end{eqnarray}
Then, for a Gaussian-random magnetic field distribution,
\begin{eqnarray}
S_\nu &=& \frac{(1+z)}{D_L^2} \frac{\sqrt{3} e^3 }{8 \pi^2 \epsilon_0 c m_e}  \\
&\times& \int_{\gamma_{\rm min}}^{\gamma_{\rm max}} N(\gamma) \left[ \int_0^\infty   \sqrt{\frac{2}{\pi}} \frac{B^3 \exp(-B^2/2a^2)}{a^3}  \bar{F} (y) dB \right]d\gamma  \nonumber
\end{eqnarray}

\begin{eqnarray}
S_\nu &=& \frac{(1+z)}{D_L^2} \frac{9e^3 }{8 \pi^2 \epsilon_0 c m_e}  \sqrt{\frac{2}{\pi}}  \\
&& \int_{\gamma_{\rm min}}^{\gamma_{\rm max}} N(\gamma) \left[ \int_0^\infty    \left( \frac{B}{B_0} \right)^3      \exp \left(-\frac{1}{6} \left(  \frac{B}{B_0} \right)^2   \right)  \bar{F} (y) dB \right]d\gamma  \nonumber
\end{eqnarray}

It therefore remains to specify the volume integrated electron energy distribution as a function of time, $N(\gamma, t)$. We do so by solving the continuity equation for the electron population within the lobes, as described in the following section.

\subsection{Volume integrated electron distribution $N(\gamma)$}  \label{app:vol_integrated_N_gamma}

Consider a radio lobe of volume V(t) with mean magnetic energy density $U_B(t) = B_0(t)^2/2 \mu_0$, into which relativistic particles are injected at a rate of $\frac{dN}{d \gamma_i d t_i} = \sigma_{\rm lobe}(\gamma_i, t_i)$. Provided that the cooling rate $\frac{d\gamma}{dt}$ is a function of time and particle energy only (assumption (ii)), the evolution of the volume integrated particle energy distribution is described by the continuity equation
\begin{equation}  
\frac{\partial N(\gamma, t)}{\partial t} + \frac{\partial}{\partial \gamma} \left(  N(\gamma, t) \frac{\partial \gamma}{\partial t}  \right) = \sigma_{\rm lobe} (\gamma, t)  
\end{equation}
The general solution to the continuity equation is
\begin{equation}   \label{eqn:N_gamma_t}
\frac{dN}{d\gamma} \left( \gamma, t  \right) = \int_{t_{\rm i, min}}^t \sigma \left( \gamma_i, t_i  \right) ~  \frac{d\gamma_i}{d\gamma} dt_i
\end{equation}
and is valid for any functional form of the cooling rate $d\gamma/dt$. Here we consider only radiative and adiabatic losses (it is straightforward to modify the following to include acceleration and/or additional loss terms). In this case we can write the cooling rate as 
\begin{equation} \label{eqn:dgamma_by_dt}
\frac{d \gamma}{dt} = -a_0 \gamma^2 U(t) -\frac{1}{3} \gamma \frac{1}{V} \frac{dV}{dt} 
\end{equation}
where, in both c.g.s. and S.I. units,
\begin{equation}
a_0 = \frac{4 \sigma_T}{3 m_e c}
\end{equation}
and $U(t)$ is the sum of the magnetic energy density and the CMB energy density. 
For arbitrary volume and magnetic field evolution, the injected Lorentz factor, $\gamma_i$, injected at time $t_i$, is related to the particle's Lorentz factor $\gamma$ at time t, via
\begin{equation} 
\gamma_i = \frac{\gamma  \left( \frac{V(t_i)}{V(t)} \right)^{-1/3} } {1 -  \frac{\gamma}{\gamma_*}}
\end{equation}
and the derivative is
\begin{equation} 
\frac{d\gamma_i}{d \gamma} = \frac{ \left( \frac{V(t_i)}{V(t)} \right)^{-1/3}  }{ \left( 1 - \frac{\gamma}{\gamma_*} \right)^2 }
\end{equation}
The completely general solution for arbitrary V(t), $B_0(t)$ and $\sigma_{\rm lobe} (\gamma_i, t_i)$ is 
\begin{equation}   \label{eqn:dN_by_dgamma_ingtegral}
 \frac{dN}{d\gamma}\left( \gamma, t  \right)  =  \int_{t_{\rm i, min}}^t   \sigma_{\rm lobe} (\gamma_i, \tau)   \left(  \frac{V(\tau)}{V(t)} \right)^{-1/3}  \left( 1 - \frac{\gamma}{\gamma_*(\tau, t)} \right)^{-2} d\tau
\end{equation}
where
\begin{equation} 
\gamma_i = \frac{\gamma  \left( \frac{V(t_i)}{V(t)} \right)^{-1/3} } {1 -  \frac{\gamma}{\gamma_*(t_i, t)}}
\end{equation}
and
\begin{equation}   \label{eqn:one_over_gamma_star_integral}
\frac{1}{\gamma_*(t_i, t)} =  \int_{t_i}^{t}  a_0 \left(  \frac{V(\tau)}{V(t)}  \right)^{-1/3} U(\tau)  d\tau 
\end{equation}
and the integration limit $t_{\rm i, min}$ is given by 
\begin{equation}  
t_{\rm i, min} = \mbox{MAX}(0, t_{\rm i, min}^*) 
\end{equation}
and $t_{\rm i, min}^*$ is given by the solution to the equation
\begin{equation} \label{eqn:t_i_min_star}
\frac{1}{\gamma} = \frac{1}{\gamma_{\rm i, max} \left( \frac{V(t_{\rm i, min}^*)}{V(t)} \right)^{1/3}}   +   \frac{1}{\gamma_*(t_{\rm i, min}^*, t)} 
\end{equation}
where $\gamma_{\rm i, max}$ is the maximum electron Lorentz factor of the particle distribution injected into the lobes. In the case of an infinite power-law, $\gamma_{\rm i, max} \rightarrow \infty$, in which case the first term on the RHS of equation \ref{eqn:t_i_min_star} is equal to zero. For injection times earlier than $t_{\rm i, min}^*$, all particles will have cooled to below the Lorentz factor $\gamma$ at time t. 

Equations \ref{eqn:dN_by_dgamma_ingtegral} - \ref{eqn:t_i_min_star} represent the generalised continuous injection model. The standard $CI_{\rm off}$ model of \citet{komissarov94}, as well as the equations of \citet{murgia99} and the model of \citet{ensslin01} can all be derived from the above expression under the relevant simplifying assumptions. 
%For example, Equation \ref{eqn:dN_by_dgamma_ingtegral} is the more general version of Equation A2 of \citet{murgia99}. 

The above expression is valid for an arbitrary particle injection spectrum $\sigma_{\rm lobe} (\gamma_i, t_i)$, and arbitrary evolution of the cooling function $d\gamma/dt$. We note that $\sigma_{\rm lobe} (\gamma_i, t_i)$ is the electron energy distribution that is ``injected" into the lobe. Typically the particles are accelerated (injected) in regions of higher pressure than the lobes (i.e. the hotspots). The relationship between the particle distribution that is injected into the lobes and the particle distribution that is accelerated in the hotspots is discussed below in Section \ref{sec:sigma_and_jet_power}.

\subsection{The relation between $\sigma_{\rm lobe} (\gamma_i, t_i)$ and Jet Power}  \label{sec:sigma_and_jet_power}

%See Manolakou \& Kirk 2002 for a discussion of this, particularly equations 20 and 21. \\

\noindent The particle injection rate into the lobe, per unit energy, is defined by 
\begin{equation}
\sigma_{\rm lobe} (\gamma_i, t_i) = \frac{dN}{d\gamma_i dt_i}
\end{equation}
We assume that the electron energy spectrum injected into the lobes can be approximated by a power law 
\begin{equation}
\sigma_{\rm lobe} (\gamma_i, t_i) = q_0 \gamma_i^{-a} \qquad \gamma_{\rm lobe, min} < \gamma_{i}  < \gamma_{\rm lobe, max}
\end{equation}
We can relate the proportionality constant $q_0$ to the jet power, if we make a number of further model assumptions. We assume that the particles are injected into a high pressure region (the hotspot) with pressure $p_{\rm hs}$ and flow into the lobe at some lower pressure $p_{\rm lobe}$. The particles lose energy as they flow from the high pressure injection site into the lobes via adiabatic expansion and radiative cooling, and may gain energy due to re-acceleration processes \citep[eg.][]{orienti12}. Due to the complexity and uncertainty surrounding the radiative cooling and re-acceleration in the head of the source, we consider only adiabatic expansion losses in this region, but note that alternative specific models may be considered \citep{blundell99, manolakou02}. 

Suppose that some (constant) fraction of the jet power, $\epsilon_e$, is converted to the internal energy of the relativistic particle population, and that the energy per particle is reduced due to adiabatic expansion losses by a factor
\begin{equation}
\frac{\gamma_{\rm lobe}}{\gamma_{\rm hs}} = \left( \frac{p_{\rm hs}}{p_{\rm lobe}}  \right)^{-1/4}
\end{equation}
The energy injection rate (into the lobes) is 
\begin{equation}
\epsilon_e Q_{\rm jet} \left( \frac{p_{\rm hs}}{p_{\rm lobe}}  \right)^{-1/4}  =  \int_{\gamma_i} \sigma_{\rm lobe} (\gamma_i, t_i) ( \gamma_i - 1) m_e c^2 d\gamma_i
\end{equation}
where the integration is over the particle distribution in the lobes, not the particle distribution injected into the hotspots. We can then write

\begin{equation} \label{eqn:q_0}
q_0 = \left( \frac{p_{\rm hs}}{p_{\rm lobe}}  \right)^{-1/4} \frac{\epsilon_e Q_{\rm jet}}{m_e c^2   f(a, \gamma_{\rm lobe, min}, \gamma_{\rm lobe, max})      } 
\end{equation}
where
\begin{equation} 
 f(a, \gamma_{\rm lobe, min}, \gamma_{\rm lobe, max}) =  \left(  \left[ \frac{\gamma_{\rm i}^{2-a}}{(2-a)}  \right]_{\gamma_{\rm lobe, min}}^{\gamma_{\rm lobe, max}}  -  \left[ \frac{\gamma_{\rm i}^{1-a}}{(1-a)}  \right]_{\gamma_{\rm lobe, min}}^{\gamma_{\rm lobe, max}} \right)
\end{equation}
and 
\begin{equation} 
\gamma_{\rm lobe, min/max} = \left( \frac{p_{\rm hs}}{p_{\rm lobe}}  \right)^{-1/4}  \gamma_{\rm hs, min/max}
\end{equation}

\section{Limiting cases of the generalised Continuous Injection Model}

\subsection{Power-law injection spectrum}

It is typically assumed that a power-law energy spectrum is injected, with
\begin{eqnarray}
\sigma(\gamma_i, t_i) &=& q_0 \gamma_i^{-a}  
\end{eqnarray}
In this case, 
\begin{equation}  
 \frac{dN}{d\gamma}\left( \gamma, t  \right)  =  q_0 \gamma^{-a} \int_{t_{\rm i, min}}^t    \left(  \frac{V(\tau)}{V(t)} \right)^{(a-1)/3}  \left( 1 - \frac{\gamma}{\gamma_*(\tau, t)} \right)^{a-2} d\tau
\end{equation}
The effects of cooling and source evolution enter via $\gamma_*$ as defined in Equation  \ref{eqn:one_over_gamma_star_integral}. 
The different limiting cases are obtained by evaluating the integral \ref{eqn:one_over_gamma_star_integral} analytically under different sets of assumptions.

\subsection{Single power law phase}

Let the source volume and magnetic field strength evolve according to power-laws 

\begin{eqnarray}
V(t) &=& V_0 \left( \frac{t}{t_0} \right)^{n_V} \\
B(t) &=& B_0 \left( \frac{t}{t_0} \right)^{n_B} \\
\Rightarrow U_B (t) &=& U_{B_0} \left( \frac{t}{t_0} \right)^{2 n_B}
\end{eqnarray}
The energy density of the cosmic microwave background is $U_{\rm CMB} = 4.2 \times 10^{-12} ~(1+z)^4$~J~m$^{-3}$.  From Equation \ref{eqn:one_over_gamma_star_integral}, it can be shown that: 
\begin{eqnarray}
\frac{1}{\gamma_*(t_1, t_2)} &=& a_0  U_B(t_2) t_2 \frac{ \left( \left( \frac{t_2}{t_1} \right)^{\frac{n_V}{3} -2 n_B  - 1}    - 1   \right)  }{\frac{n_V}{3} -2 n_B  - 1}  \nonumber \\
&+&  a_0 U_{\rm CMB} t_2 \frac{ \left( \left( \frac{t_2}{t_1} \right)^{\frac{n_V}{3} - 1}    - 1   \right)  }{\frac{n_V}{3} - 1}    
\end{eqnarray}
This is equivalent to equation 7 of \citet{ensslin01}, and equation 7 of \citet{kda97}. 

\subsection{Multiple power law phases}

Suppose we have K power-law phases of evolution, bounded by the times $(t_0, t_1, t_2, ..., t_K)$ for phases (1, 2, ..., K) and each $j^{\rm th}$ phase has its own $n_{V_j}$ and $n_{B_j}$. Then for each $j^{\rm th}$ phase we can define $\gamma_{*,j}$ as

\begin{equation}  
\frac{1}{\gamma_{*,j}} =  \frac{1}{\gamma_*(t_{j-1}, t_j)} =  \int_{t_{j-1}}^{t_j}  a_0 \left(  \frac{V(\tau)}{V(t_j)}  \right)^{-1/3} U(\tau)  d\tau
\end{equation}
The total effect of all of the phases is given by
\begin{eqnarray}  
\frac{1}{\gamma_*(t_{0}, t_K)} &=&  \int_{t_{0}}^{t_K}  a_0 \left(  \frac{V(\tau)}{V(t_K)}  \right)^{-1/3} U(\tau)  d\tau \\
&=&  \int_{t_{0}}^{t_1}  a_0 \left(  \frac{V(\tau)}{V(t_K)}  \right)^{-1/3} U(\tau)  d\tau   \\ 
&& +  \int_{t_{1}}^{t_2}  a_0 \left(  \frac{V(\tau)}{V(t_K)}  \right)^{-1/3} U(\tau)  d\tau  \nonumber  \\ 
&& + ... \nonumber \\
&& + \int_{t_{K-1}}^{t_K}  a_0 \left(  \frac{V(\tau)}{V(t_K)}  \right)^{-1/3} U(\tau)  d\tau \nonumber  \\
&=& \sum_{j=1}^K \frac{1}{\gamma_{*, j}} \left( \frac{V(t_j)}{V(t_K)} \right)^{-1/3} 
\end{eqnarray}
The above is a slightly different, but equivalent form to that given by \citet{ensslin01}. 

Suppose that injection occurs only within the first of several phases, i.e. prior to $t_1$, then
\begin{equation}  
\frac{dN}{d\gamma}\left( \gamma, t  \right)  =  \int_{t_{\rm i, min}}^{t_1}   q_0 \gamma^{-a}  \left(  \frac{V(\tau)}{V(t)} \right)^{(a-1)/3}  \left( 1 - \frac{\gamma}{\gamma_*(\tau, t)} \right)^{a-2} d\tau 
\end{equation}
where $\gamma_*(\tau, t)$ is given by the above expression, with $t_0 = \tau$ and $t_K = t$. 

To test the formalism described in this section, we compared the output with a numerical solution of the continuity equation using a finite volume method with piecewise linear reconstruction and a van Leer flux limiter. The tests demonstrate that the above formalism is correct, and provides much faster evaluation than is possible with the finite volume method while avoiding the effects of numerical diffusion.

\section{The angle-averaged Synchrotron function $\bar{F}(y)$ and it's evaluation}  \label{app:F_bar_y}

For an isotropic distribution of magnetic field orientation, the distribution of angle $p_\Theta = \frac{1}{2} \sin \Theta$. We can then write the inner integral
\begin{equation}
\bar{F}(y) = \int_0^\pi \sin \Theta F(x)  p_\Theta  d\Theta  = \int_0^{\pi/2} F(x)  \sin^2 \Theta  d\Theta
\end{equation}
where $F(x)$ is defined in equation \ref{eqn:j_nu_appendix_eqn}. Let 
\begin{equation}
y = \frac{\nu}{\nu_c} \sin \Theta = x \sin \Theta = \left( \frac{4 \pi \nu}{3 \Omega_0 \gamma^2} \right)
\end{equation}
Then we can write
\begin{eqnarray}
\bar{F} (y) &=&  \int_0^{\pi/2} \sin^2 \Theta~  F\left( \frac{y}{\sin \Theta} \right) d \Theta \nonumber \\
&=&  \int_0^{\pi/2} \left[  \frac{y}{\sin \Theta}  \int_{ \frac{y}{\sin \Theta} }^\infty K_{5/3} (z)~ dz \right]  \sin^2 \Theta~  d \Theta \nonumber \\
&=&  y \int_y^{\infty}   \left(  1 - \frac{y^2}{t^2} \right)^{1/2}    K_{5/3} (t)~ dt    \label{eqn:F_bar_proper}
\end{eqnarray}
where $K_{5/3}(z)$ is the modified Bessel function of order 5/3. \citet{crusius86} obtain an expression of $\bar{F}(y)$ in terms of Whittaker functions:

\begin{equation}
\bar{F} (y) = \frac{1}{2} \pi y \left[ W_{0, \frac{4}{3}}(y)~ W_{0, \frac{1}{3}}(y) - W_{\frac{1}{2}, \frac{5}{6}}(y)~W_{-\frac{1}{2}, \frac{5}{6}}(y)    \right]
\end{equation}
where $W_{\lambda, \mu}(y)$ denotes Whittaker's function \citep[][pg. 505]{abramowitz70}. 
%Note that Whittaker's function can be evaluated in Python using the function  \textit{\textbf{whitw}} from the module \textit{\textbf{mpmath}}. 

\citet{ensslin99} give an approximation to the function $\bar{F}(y)$ to within a few percent
\begin{equation}
\bar{F} (y) = \frac{2^{2/3}}{\Gamma(11/6) \left( \frac{\pi}{3} \right)^{3/2} y^{1/3} } \exp \left( -\frac{11}{8} y^{7/8} \right)
\end{equation}
However, such an approximation is not necessary when spline interpolation can provide an efficient and accurate computational solution. For small values of $y$ ($y \lesssim 0.01$), an asymptotic form is available (G. Bicknell, private communication): 
\begin{equation}
\bar{F} (y) \approx c_1 ~ y^{1/3} + c_2~ y + c_3 ~ y^{7/3} + c_4~ y^{3}   \> \qquad y << 1
\end{equation}
where 
\begin{itemize}
    \item $c_1=1.808418021$ 
    \item $c_2=-1.813799364$ 
    \item $c_3=0.8476959474$
    \item $c_4=-0.510131$
\end{itemize}

%\noindent Due to the exponential drop of $\bar{F}(y)$ at large values of $y$, it is appropriate to set $\bar{F}(y) = 0$ for $y > 20$. 

%%%%%%%%%%%%%%%%%%%%%%%%%%%%%%%%%%%%%%%%%%%%%%%%%%

% Don't change these lines
\bsp	% typesetting comment
\label{lastpage}
\end{document}